\documentclass{article}%
\usepackage{amsmath}
\usepackage{amsfonts}
\usepackage{amssymb}
\usepackage{graphicx}%
\providecommand{\U}[1]{\protect\rule{.1in}{.1in}}
\newtheorem{theorem}{Theorem}

\newtheorem{proposition}[theorem]{Proposition}
\newtheorem{remark}[theorem]{Remark}

\begin{document}

\title{From the Photon to Maxwell Equation. Ponderations on the Concept of
Photon\ Localizability and Photon Trajectory in a de
Broglie-Bohm\ Interpretation of Quantum Mechanics}
\author{Waldyr A. Rodrigues Jr.\\Institute of Mathematics Statistics and Scientific Computation\\IMECC-UNICAMP\\walrod@ime.unicamp.br or walrod@mpc.com.br}
\maketitle
\tableofcontents

\begin{abstract}
In this paper using the Clifford bundle formalism we show how starting from
the photon concept and its relativistic Hamilton-Jacobi equation (HJE) we
immediately get (with a simple hypothesis concerning the form of the photon
canonical momentum) Maxwell equation (ME) satisfied by a null $2$-form field
$\boldsymbol{F}$ which is a plane wave solution (PWS) of ME. Moreover, we show
how introducing a potential $1$-form $\boldsymbol{A}$ such that
$\boldsymbol{F}=d\boldsymbol{A}$ we can see how a duality rotation changed in
a \ spatial rotation transformation besides showing how $\mathrm{i}=\sqrt{-1}$
enters Maxwell theory thus permitting the writing of a representative of ME as
Schr\"{o}dinger like equation which plays a key role in answering one of the
main questions addressed in this paper, namely: is there any sense in talking
about photon trajectories in de Broglie-Bohm like theories? To this end we
investigate first the nature of the energy-momentum \emph{extensor} field of
the Maxwell field $\mathbf{T}(n)$ in some special situations showing that for
some of those cases $\mathbf{T}_{0}=\mathbf{T}(\boldsymbol{\gamma}_{0})$ even
cannot describe the flow of energy. A proposed solution is offered. We also
prove that there exists a pulse reshaping phenomenon even in vacuum. Finally
we discuss the Schr\"{o}dinger equation for a photon that follows from quantum
field theory and investigate solutions that some authors think imply in photon
localization. We discuss if such an idea is meaningful. Moreover, we show that
for such solutions it is possible (once we accept that photon wave functions
are extended in the space and also in the time domains) to derive a
generalized HJE containing a quantum potential and which may lead to non
lightlike photon trajectories in free space. We briefly discuss our findings
in relation to a recent experiment.

\end{abstract}

\section{Introduction}

In \cite{rs1} we showed that using the Clifford bundle formalism it is
possible to obtain a \emph{first order} HJE directly from the classical
relativistic (quadratic) HJE for a charged spin $1/2$ particle in interaction
with an external electromagnetic field. It is then shown that the first order
HJE is equivalent to a Dirac-Hestenes equation satisfied by a special class of
Dirac-Hestenes spinor fields (DHSF) that are characterized by having its
Takabayashi angle function equal to $0$ or $\pi.$It is also shown that the
Dirac-Hestenes equation satisfied by general by a general DHSF\ produces a
generalized relativistic Hamilton-Jacobi equation implying (if this is a licit
implication) that the particle follows a trajectory resulting from the action
of the external electromagnetic field \ plus the action of a new
potential\footnote{It is to emphasized here that the new quantum potential
differs from the ones in the original de Broglie-Bohm theories.} which seems
to be the correct relativistic quantum potential. It is very important to
observe that all these results have been obtained without the use of the
quantum mechanics formalism. Thus, they suggest of course, a de Broglie-Bohm
interpretation of the quantum formalism \footnote{Our statement does not mean
that we endorse this theory as the correct interpretation of the quantum
formalism \ Indeed, we leave clear here that we think that the only coherent
interpretation of quantum phenomena is the one given by quantum field
theory.}. Some aspects of the results obtained in \cite{rs1} and criticisms to
some attempts to build a de Broglie-Bohm theory for fermions has been
discussed in \cite{mrs}.

Motivated from the above results here (with the aid of the Clifford bundle
formalism) we show in Section 2 how starting from the photon concept (a zero
mass particle) and its classical relativistic HJE we immediately get the free
Maxwell equation (ME) once we postulate that $\boldsymbol{\partial}S=\frac
{1}{2}\boldsymbol{F}\boldsymbol{\gamma}^{0}\boldsymbol{F}$ satisfied by a null
$2$-form field $\boldsymbol{F=}$ $\mathcal{F}_{0}e^{\boldsymbol{\gamma}^{5}S}%
$, where $S$ is the classical action and $\boldsymbol{\gamma}^{5}$ is the
volume element of Minkowski\footnote{Note that $\mathbf{T}(n)$ differs from
the \emph{canonical} energy-momentum extensor of the electromagnetic field,
whose expression is recalled in Appendix B.} spacetime. After getting ME,
namely $\boldsymbol{\partial F}=0$, where $\boldsymbol{\partial}$ is the Dirac
operator (see Appendix A) we show that this equation automatically produces a
conserved energy momentum extensor field $\mathbf{T}:\sec%
{\textstyle\bigwedge\nolimits^{1}}
T^{\ast}M\hookrightarrow\sec\mathcal{C\ell(}M,\mathtt{g})\ni$ $n\mapsto
\mathbf{T}(n)\in$.$\sec%
{\textstyle\bigwedge\nolimits^{1}}
T^{\ast}M\hookrightarrow\sec\mathcal{C\ell(}M,\mathtt{g})$. The object
$\mathbf{T}^{0}=\mathbf{T}(\boldsymbol{\gamma}^{0})=-\frac{1}{2}%
F\boldsymbol{\gamma}^{0}F$ is one of the (symmetrical) energy-momentum
$1$-form fields of the Maxwell field which plays a key role in our
considerations concerning the issues about if there is any meaning in the
concept of photon trajectories, the localizability of these objects and the
description of the energy transport and its velocity for an arbitrary
electromagnetic field configuration $\boldsymbol{F}$ (solution of ME).\ In
Section 3 we recall the Hertz potential method to generate solutions of the
free ME, recall that it it possess besides the well known luminal (also called
null fields, since characterized for satisfying. $\boldsymbol{F}^{2}=0$)
extraordinary free boundary solutions describing hypothetical subluminal and
superluminal electromagnetic field configurations (characterized by having
$\boldsymbol{F}^{2}\neq0$).\ In Section 4 we explicitly discuss the case of
the superluminal electromagnetic $X$-wave, and from that solution we construct
a\emph{ finite aperture approximation }to this wave\footnote{Which can be
launched by a special antenna.}, which has finite energy, and for which its
front and rear travels with speed\footnote{In this paper we use natural unitis
where the speed of light in vacuum $c$ has the numerical value $1$. Also, $1$
is the numerical value of the Planck constant $\hbar$.} $v=c=1$, whereas its
peak travels for a \emph{while} with superluminal speed. We show that this
notable situation occurs due to the \emph{reshaping phenomenon} (occurring in
vacuum) which we careful discuss. We discuss the nature of $\mathbf{T}^{0}$
for such finite aperture approximation showing that $\mathbf{T}^{0}$ is a
timelike $1$-form field and thus cannot describe the propagation of the
energy. A eventual solution for this dilemma is proposed. We also analyze in
Section 5 the nature of $\mathbf{T}^{0}$ for the case of a static electric
plus magnetic field configuration where $\mathbf{T}^{0}$ is non null although
nothing \ \textquotedblleft material\textquotedblright\ l is seem to be in
motion. In Section 6 we show how introducing a potential $1$-form
$\boldsymbol{A}$ such that $\boldsymbol{F}=d\boldsymbol{A}$ the \emph{duality
rotation} $e^{\boldsymbol{\gamma}^{5}S}$ metamorphosis in a rotation
transformation thus showing how $i=\sqrt{-1}$ enters Maxwell theory. In
Section 6 we show also how to trivially obtain within our formalism a
Schr\"{o}dinger form\footnote{Such a form is the one used, e.g., in
\cite{jajan} to discuss what thee authors think to be the maximum possible
localization of photons.} of ME originally discovered by Riemann and
rediscovered by many other authors.

Until Section 6 the analysis is almost classical. So, in Section 7 we
investigate the quantum Schr\"{o}dinger equation for the photon which follows
from quantum field theory, concentrating analysis on the solution, one given
in \cite{bia} and the other in \cite{sbz}. In \cite{bia} the author claims to
\ produce a photon localizability better than earlier attempts. We discuss
such a claim and explicitly show that this statement if accepted implies in
very odd facts concerning the world we live in. Nevertheless if we are
prepared to accept such odd facts and if we suppose that it is meaningful to
talk about photon trajectories than we can derive a \ more general HJE for the
photon more general than the one that has been used to derive from the photon
concept the ME. This new equation implies that photon trajectories are not
lightlike geodesics of Minkowski spacetime (they can be timelike) This
produces eventually surreal trajectories, and we will study more this issue in
another publication. We also recall in Section 7 the focus wave mode
representation of the photon described in \cite{sbz} and comment on the
possible photon trajectories associated with this solution. Section 7 ends
with pertinent comments concerning some statements\ in \cite{fh2016}
concerning their Bohm like theory of photon trajectories and the results of a
recent experiment \cite{mrf} that claims to observe photon trajectories.
Conclusions are in Section 8.

The paper has several appendices. In Appendix A we show how to nicely obtain
the \textquotedblleft vector form\textquotedblright\ of Maxwell equations from
the single ME written in the Clifford bundle. The importance of this
derivation is the obtain a representation of ME that leads directly to the one
originally obtained by Riemann. In Appendix B we recall how to obtain within
the Clifford bundle formalism the symmetric energy-momentum extensor
$\mathbf{T}(n)$ and the angular momentum extensor $\mathbf{J}$%
\textbf{$^{\dagger}$}$(n)$ of the electromagnetic field. We also recall the
form of the Poincar\'{e} invariants and recall that the spin of the
electromagnetic field can only be obtained form the Lagrangian formalism
through the canonical energy-momentum extensor \cite{rc2016}.

\section{From Photon to Maxwell Equation}

In this paper we suppose that all phenomena take place in Minkowski spacetime,
i.e., the structure $(M\simeq\mathbb{R}^{4},\boldsymbol{g},\boldsymbol{D}%
,\tau_{\boldsymbol{g}},\uparrow)$, where $(M\simeq\mathbb{R}^{4}%
,\boldsymbol{g},\tau_{\boldsymbol{g}},\uparrow)$) is an oriented Lorentzian
manifold (oriented by $\tau_{\boldsymbol{g}}\in\sec%
{\textstyle\bigwedge\nolimits^{4}}
T^{\ast}M$) and time oriented by $\uparrow$ (details in \cite{rc2016}), and
$\boldsymbol{D}$ is the Levi-Civita connection of $\boldsymbol{g}$.$\in\sec
T_{2}^{0}M$, a Lorentzian metric of signature $(1,3)$. We shall use in this
paper only coordinates in Einstein-Lorentz-Poincar\'{e} gauge $\{x^{\mu}\}$,
for which\footnote{The matrix with entries $\eta_{\mu\nu}$ is the diagonal
matrix \textrm{diag}$(1,-1,-1,-1)$.} $\boldsymbol{g}=\eta_{\mu\nu
}\boldsymbol{\gamma}^{\mu}\otimes\boldsymbol{\gamma}^{\nu}$%
,with$\{\boldsymbol{\gamma}^{\mu}=dx^{\mu}\}$ the dual basis of $\{e_{\mu
}=\partial_{\mu}\}$. We denote by $\mathtt{g}=\eta^{\mu\nu}%
\boldsymbol{\partial}_{\mu}\otimes\boldsymbol{\partial}_{\nu}$ the metric on
$T^{\ast}M$, such that $\eta^{\mu\rho}\eta_{\rho\nu}=\delta_{\nu}^{\mu}$. We
use also in the text the cobasis $\{\boldsymbol{\gamma}_{\mu}\}$ which is the
reciprocal basis of $\{\boldsymbol{\gamma}^{\mu}\}$, i.e., $\mathtt{g}%
(\boldsymbol{\gamma}^{\mu},\boldsymbol{\gamma}_{\nu})=\delta_{\nu}^{\mu}$.
Moreover, we suppose that all fields involved are sections of $%
{\textstyle\bigwedge}
T^{\ast}M\hookrightarrow\mathcal{C\ell(}M,\mathtt{g})$, where $%
{\textstyle\bigwedge}
T^{\ast}M=%
{\textstyle\bigoplus\nolimits_{i=0}^{4}}
$ $%
{\textstyle\bigwedge^{i}}
T^{\ast}M$ is the bundle of non homogenous differential forms and
$\mathcal{C\ell(}M,\mathtt{g})$ is the Clifford bundle of differential forms.
The typical fiber of $\mathcal{C\ell(}M,\mathtt{g})$ is $\mathbb{R}_{1,3}$,
the so-called spacetime algebra. The even subalgebra of $\mathcal{C\ell
(}M,\mathtt{g})$ is denoted $\mathcal{C\ell}^{0}\mathcal{(}M,\mathtt{g})$ and
its typical fiber $\mathbb{R}_{3,0}$ the so-called Pauli algebra. Moreover
$\boldsymbol{\partial}=\boldsymbol{\gamma}^{\mu}\partial_{\mu}$ is the Dirac
operator acting on sections of the Clifford bundle.

So to start we suppose that photons follows in Minkowski spacetime light like
curves. Let then be $\sigma:\mathbb{R\rightarrow}M$, with $\boldsymbol{g}%
(\sigma_{\ast},\sigma_{\ast})=0$ the worldline of a given photon. Now, imagine
a vector field $V\in\sec TM$ such that $\left.  V\right\vert _{\sigma}%
=\sigma_{\ast}$ and $\boldsymbol{g}(V,V)=0$. Moreover write $p=\alpha
\sigma_{\ast}$ (with $\alpha\in\mathbb{R}$, a constant) the momentum of the
photon and put $P=\alpha V$. Also, define the $1$-form fields $\boldsymbol{V}%
=\boldsymbol{g}(V\mathbf{,~})$ and $\boldsymbol{P}=\boldsymbol{g}%
(P\mathbf{,~})$. Of course, $\boldsymbol{P}$ must be understood in what
follows as a \emph{density} of momentum

Next we suppose that $\boldsymbol{P}$ is exact, i.e., there exists a function
$S\in\sec%
{\textstyle\bigwedge^{0}}
T^{\ast}M\hookrightarrow\sec\mathcal{C\ell(}M,\mathtt{g})$ such that%
\begin{equation}
\boldsymbol{P}:=-dS=-\boldsymbol{\partial}S\in\sec%
{\textstyle\bigwedge\nolimits^{1}}
T^{\ast}M\hookrightarrow\sec\mathcal{C\ell(}M,\mathtt{g}). \label{0}%
\end{equation}

Of course, we have
\begin{equation}
\mathtt{g}(\boldsymbol{\partial}S,\boldsymbol{\partial}%
S)=(\boldsymbol{\partial}S)^{2}=\boldsymbol{P}^{2}=0. \label{1}%
\end{equation}
where
\begin{equation}
(\boldsymbol{\partial}S)^{2}=\boldsymbol{\partial}S\cdot\boldsymbol{\partial
}S=0 \label{1n}%
\end{equation}
is the classical Hamilton-Jacobi equation for a free photon.

We now introduce a $\boldsymbol{F}\in\sec%
{\textstyle\bigwedge^{2}}
T^{\ast}M\hookrightarrow\sec\mathcal{C\ell(}M,\mathtt{g})$ such that%
\begin{equation}
\boldsymbol{P}=\mu\boldsymbol{F\gamma}^{0}\boldsymbol{F} \label{2n}%
\end{equation}
where $\mu$ is a constant whose nature will be investigated below.

Now, write%

\begin{equation}
\boldsymbol{F}=\mathcal{F}_{0}e^{\boldsymbol{\gamma}^{5}S}\in\sec%
{\textstyle\bigwedge\nolimits^{2}}
T^{\ast}M\hookrightarrow\sec\mathcal{C\ell(}M,\mathtt{g}) \label{2}%
\end{equation}
where $\mathcal{F}_{0}\in\sec%
{\textstyle\bigwedge\nolimits^{2}}
T^{\ast}M\hookrightarrow\sec\mathcal{C\ell(}M,\mathtt{g})$ is a constant
biform. Since $\boldsymbol{P}^{2}=0$ we need $\boldsymbol{F}^{2}=0$ and since
it is also $\boldsymbol{P}=\mu\mathcal{F}_{0}\boldsymbol{\gamma}%
^{0}\mathcal{F}_{0}e^{2\boldsymbol{\gamma}^{5}S}$\ we need also to impose that
$\mathcal{F}_{0}^{2}=0$.

Now, free photons are supposed to follows null geodesics of the Minkowski
spacetime, so this implies that $\boldsymbol{P}$ is a constant $1$-form field.

From Eq.(\ref{2}) we get
\begin{equation}
-\boldsymbol{\gamma}^{5}\boldsymbol{\partial F}=\boldsymbol{\partial
}S\boldsymbol{F}=\boldsymbol{PF} \label{3n}%
\end{equation}
and using Eq.(\ref{2n}) in Eq.(\ref{3n}) with $\boldsymbol{F}^{2}=0$ it
follows that\footnote{The form of Maxwell equation given by Eq.(\ref{4n}) was
first obtained by M. Riez \cite{riez} as a simple translation using Clifford
algebras of the Maxwell equations written in the vector formalism. It has been
divulged by Hestenesin his wonderful book \cite{hestenes} Take notice that
here the equation has been obtained from the photon concept.}%
\begin{equation}
\boldsymbol{\partial F}=0, \label{4n}%
\end{equation}
i.e., we have proved that under the above assumptions the Hamilton-Jacobi
equation for the photon (Eq.(\ref{1n})) implies the validity of \emph{free
Maxwell equation } for a special kind of fields, the ones where
$\boldsymbol{F}^{2}=0$ and $\mathcal{F}_{0}$ is a constant $2$-form field.
Also, since%
\begin{equation}
\boldsymbol{\partial F}=\boldsymbol{\gamma}^{5}\boldsymbol{\partial
}S\boldsymbol{F} \label{4nnn}%
\end{equation}
we have on multiplying both members by $\boldsymbol{\partial}S$ and taking
into account Eq.(\ref{4n}) that
\begin{equation}
0=-\boldsymbol{\gamma}^{5}\boldsymbol{\partial}S\boldsymbol{\partial
F}=\boldsymbol{\partial}S\cdot\boldsymbol{\partial}S\boldsymbol{F} \label{4NA}%
\end{equation}
\noindent i.e., $\boldsymbol{\partial}S\cdot\boldsymbol{\partial}S=0$. Thus we
can state that under the above conditions the Hamilton-Jacobi equation for a
photon and the free Maxwell equation are equivalent.\medskip

Moreover, we have discovered a momentum operator\footnote{In \cite{rc2016} the
symbol \ $\symbol{94}$ has been used to denote the main involution in Clifford
algebras. In this paper a hat over a symbol will denote an operator.}%
\[
\boldsymbol{\hat{P}}\mathbf{:=-}\boldsymbol{\gamma}^{5}\boldsymbol{\partial}%
\]
\ acting on the sections of the Clifford bundle such that $\boldsymbol{\hat
{P}F}=\boldsymbol{PF}$.

\subsection{Nature of $\boldsymbol{F}$ for a Photon Satisfying Eq.(\ref{2})}

With Eq.(\ref{4n}) the Eq.(\ref{3n}) also implies that
\begin{equation}
\boldsymbol{PF}=0 \label{4na}%
\end{equation}
Suppose that the photon moves in the $z$-direction. Then, if $\omega$ is the
frequency\footnote{Recall that we re using natural unities where the numerical
values of the speed of light and Planck constant is equal to $1$.} of the
photon the density of momentum of its \textquotedblleft wave\textquotedblright%
\ is
\begin{equation}
\boldsymbol{P}=\omega(\boldsymbol{\gamma}^{0}+\boldsymbol{\gamma}%
^{3})e^{2\boldsymbol{\gamma}^{5}\boldsymbol{P}_{\mu}x^{\mu}}. \label{4nn}%
\end{equation}
We see that Eq.(\ref{4na}) with a $\boldsymbol{F}$ such $\boldsymbol{F}^{2}=0$
is satisfied if we put, e.g.,
\begin{equation}
\boldsymbol{F}=\left(  \boldsymbol{\gamma}^{0}\boldsymbol{\gamma}%
^{1}-\boldsymbol{\gamma}^{1}\boldsymbol{\gamma}^{3}\right)
e^{\boldsymbol{\gamma}^{5}\boldsymbol{P}_{\mu}x^{\mu}} \label{5a}%
\end{equation}
\noindent i.e., if $\boldsymbol{F}$ is a plane wave with electric and magnetic
fields orthogonal to the propagation direction.

Moreover using Eq.(\ref{5a}) in Eq.(\ref{2n}) we get:%
\begin{equation}
\boldsymbol{P}=-2\mu(\boldsymbol{\gamma}^{0}+\boldsymbol{\gamma}%
^{3})e^{2\boldsymbol{\gamma}^{5}\boldsymbol{P}_{\mu}x^{\mu}} \label{5b}%
\end{equation}
from where we infer that $\mu=-\frac{1}{2}\omega$.

\subsection{The Conserved Energy-Momentum Extensor of the Electromagnetic
Field}

Let $n$ be a constant $1$-form field such that $n^{2}=1$ or $n^{2}=-1$. The
conserved symmetrical energy-momentum extensor of the electromagnetic field is
the mapping (see Appendix B)%
\begin{equation}
\mathbf{T}:n\mapsto\mathbf{T}(n)=-\frac{1}{2}\boldsymbol{F}n\boldsymbol{F}%
\in\sec%
{\textstyle\bigwedge\nolimits^{1}}
T^{\ast}M\hookrightarrow\sec\mathcal{C\ell(}M,\mathtt{g}) \label{5c}%
\end{equation}
and we see that
\begin{equation}
\boldsymbol{P}=\mathbf{T}(\boldsymbol{\gamma}^{0}):=\mathbf{T}^{0}=T_{\mu}%
^{0}\boldsymbol{\gamma}^{\mu}, \label{5D}%
\end{equation}
i.e., the worldlines of the photons are the integral lines of the vector field
$\mathtt{g}(\mathbf{T}^{0},~)$ which is as well known the flux of
energy-momentum. In classical electromagnetic theory the \emph{energy
momentum} of an electromagnetic field configuration is the $1$-form (not a
$1$-form field)\footnote{Thr $\mathcal{E}^{\mu}$ are $1$-forms in a vector
space $\mathbb{R}^{4}$ equipped with a Minkowski scalar product. Details in
\cite{rs2016}.}%
\begin{equation}
\mathcal{P}=\mathcal{E}^{\mu}P_{\mu}=\mathcal{E}^{\mu}\int\star\mathbf{T}%
(\boldsymbol{\gamma}_{\mu}), \label{5e}%
\end{equation}
\ \noindent which gives divergent values for the $P_{\mu}$ for a free photon.
This shows that although we arrived from the photon concept to the free
Maxwell equation, the classical theory cannot deals with the photon concept
and as well known a solution requires quantum field where the energy density
of a field configuration in a general $n$-photon state is defined by normal
order of the creation and destruction operators. We will comment more on this
issue below.

\section{Extraordinary Solutions of the Free Maxwell Equation}

In \cite{rs1,mrs} we arrived at the conclusion that experimental reasons
(explanation of the hydrogen atom spectrum)\ forced us to suppose that the
Dirac-Hestenes equation satisfied by general spinor fields is to be taken as a
fundamental equation\footnote{In the sense of relativistic quantum mechanics.
Of course, as well known, this is still not enough in order to have a coherent
theory of interacting charged fermions fields with the quantized
electromagnetic field, for which relativistic \emph{quantum }field theory is
necessary.} despite the fact that there it was shown that the (linearized)
relativistic Hamilton-Jacobi equation is equivalent to the Dirac-Hestenes
equation satisfied by a special class of Dirac-Hestenes spinor fields, the
ones that have a null Takabayashi angle function.

So, it is appropriate to recall here that long ago it was found that
\cite{or1998,ortx2001,or2001,rm1996,rl1997,rv1997,rtx2001} Maxwell equation
$\boldsymbol{\partial F}=0$ as a fundamental equation has extraordinary
solutions\footnote{Solutions that have been called in \cite{rl1997}
undistorted progressive waves (UPWs).Take notice that all those extraordinary
solutions cannot be realized as physical fields. However, finite aperture
approximations to these solutions have been launched by special antennas and
for a while show peaks travelling at superluminal (or subluminal) speeds while
the fronts travel always at the light speed. Thus this phenomena does not
implies in any violation of the principle of relativity. All theses issues are
discussed with details in the references quoted above.
\par
{}}. Such solutions are characterized by $\boldsymbol{F}^{2}\neq0$ and
describe to \emph{subluminal} and \emph{superluminal }\ propagating field
configurations. We now briefly recall how those extraordinary solutions were
found .

The idea \cite{rv1997,rl1997,rc2016} is to introduce an object
$\boldsymbol{\Pi}\in\sec%
{\textstyle\bigwedge\nolimits^{2}}
T^{\ast}M\hookrightarrow\sec\mathcal{C\ell(}M,\mathtt{g})$, called the\emph{
Hertz potential},.which is supposed to satisfy the wave equation
$\boldsymbol{\partial}^{2}$$\boldsymbol{\Pi}=0$. It generates the
electromagnetic potential $\boldsymbol{A}\in\sec%
{\textstyle\bigwedge\nolimits^{1}}
T^{\ast}M\hookrightarrow\sec\mathcal{C\ell(}M,\mathtt{g})$ for which we may
evaluate $\boldsymbol{F}=d\boldsymbol{A}$. through the definition
\begin{equation}
\boldsymbol{A}:=-\delta\boldsymbol{\Pi}=\boldsymbol{\partial}\lrcorner
\boldsymbol{\Pi}. \label{7n}%
\end{equation}

Before proceeding recall that $\delta\boldsymbol{A}=0$, i.e., $\boldsymbol{A}$
is a potential (for $\boldsymbol{F}$) in Lorenz gauge.

Now, we have
\begin{equation}
\boldsymbol{\partial A}=(d-\delta)(-\delta\boldsymbol{\Pi})=d\boldsymbol{A}%
=\boldsymbol{F}=-d\delta\boldsymbol{\Pi}. \label{8}%
\end{equation}

Thus,%
\begin{equation}
\boldsymbol{\partial F}=-(d-\delta)d\delta\boldsymbol{\Pi}=\delta
d\delta\boldsymbol{\Pi} \label{9}%
\end{equation}
and since
\begin{equation}
\boldsymbol{\partial}^{2}\boldsymbol{\Pi}=0=-(d\delta+\delta d)\boldsymbol{\Pi
}=0 \label{10}%
\end{equation}
it follows that
\begin{equation}
d\delta\boldsymbol{\Pi}=-\delta d\boldsymbol{\Pi} \label{11}%
\end{equation}

Thus substituting Eq.(\ref{11}) in Eq.(\ref{9}) we get that
\begin{equation}
\boldsymbol{\partial F}=0. \label{12}%
\end{equation}

Now, it is a well fact that the homogeneous wave equation for $\phi\in\sec%
{\textstyle\bigwedge\nolimits^{0}}
T^{\ast}M\hookrightarrow\sec\mathcal{C\ell(}M,\mathtt{g})$, i.e.,
\begin{equation}
\boldsymbol{\partial}^{2}\phi=0 \label{13}%
\end{equation}
has an infinity number of subluminal and superluminal free boundary solutions
\cite{rm1996,rl1997}. So, if, e.g., $\phi$ is a superluminal solution of
Eq.(\ref{13}) then, e.g. putting $\boldsymbol{\Pi}=\phi\boldsymbol{\gamma}%
^{1}\boldsymbol{\gamma}^{2}$ it is obvious that $\boldsymbol{\Pi}$ is a
superluminal solution of $\boldsymbol{\partial}^{2}\boldsymbol{\Pi}$ and thus
also a superluminal solution of Maxwell equation.

\begin{remark}
It is important to emphasize here in order to avoid any possible
misunderstanding that\ most of the extraordinary superluminal and subluminal
solutions of Maxwell equations have, as is the case of plane waves of infinity
energy besides having no fronts and rears. Moreover, it is a surprising fact
that we can solve Maxwell equation with initial conditions for a field
configuration moving in the $z$-direction bounded between say $z=-d$ to $z=d$
but with noncompact support in the $xy$ plane such that the front and rear of
the field configurations moves at superluminal speed. This result does not
violate well known results on the Cauchy problem for hyperbolic differential
equations \emph{\cite{vla}} because the initial field configuration is not of
compact support. Details may be found in \emph{\cite{or2001}}. Also, it is
opportune to mention that finite aperture approximations to superluminal
\emph{(}and also subluminal\emph{)} solutions of Maxwell equation
\emph{(}which, of course, have finite energy\emph{)} have been launched \ in
several experiments and it has been observed that their peaks travel for a
while with superluminal (subluminal) speed. The phenomena ends when the peak
meets the front of the wave with always travel at light speed. Some details on
this fascinating subject may be found in \emph{\cite{or1998,rtx2001,ortx2001}%
}. An explicit example is given in Section 4.
\end{remark}

\subsection{$\mathbf{T}^{0}$ Expressed in Terms of the Hertz Potential}

Recalling Eq.(\ref{5b}) and Eq.(\ref{8}) and recalling that the Hertz
potential satisfy the wave equation (Eq.(\ref{10})) we immediately get%
\begin{equation}
\mathbf{T}^{0}=-\frac{1}{2}\boldsymbol{F\gamma}^{0}\boldsymbol{F}=-\frac{1}%
{2}d\delta\mathbf{\Pi}\boldsymbol{\gamma}^{0}d\delta\mathbf{\Pi=}-\frac{1}%
{2}\delta d\mathbf{\Pi}\boldsymbol{\gamma}^{0}\delta d\mathbf{\Pi} \label{T1}%
\end{equation}

Eq.(\ref{T1}) incidentally shows that the Hertz potential cannot be an exact
form since it that was the case we would have $\mathbf{T}^{0}=0$. Moreover,
take into account that forn any solution such that any solution of ME such
that $\boldsymbol{F}^{2}=\boldsymbol{F\lrcorner F+F}\wedge\boldsymbol{F}\neq0$
we have (taking into account that $\boldsymbol{F}\wedge\boldsymbol{F}%
=\boldsymbol{\gamma}^{5}L$ where $L$ is a $0$-form field that
\begin{equation}
\mathbf{T}_{0}\cdot\mathbf{T}_{0}=\frac{1}{4}[(\boldsymbol{F\lrcorner F}%
)^{2}-(\boldsymbol{F}\wedge\boldsymbol{F})^{2}]>0, \label{T2}%
\end{equation}
i.e., $\mathbf{T}_{0}$ is a timelike $1$-form field.

\section{The Superluminal Electromagnetic $X$-Wave. Pulse Reshaping and the
Character of $\mathbf{T}^{0}$}

\subsection{The $X$-Wave}

Consider first the HWE for $\Phi$ in free space and let be $\widetilde{\Phi
}(\omega,\mathbf{k})$ be the Fourier transform of $\Phi(t,\mathbf{x})$, i.e.,%
\begin{align}
\widetilde{\Phi}(\omega,\mathbf{k})  &  ={\int_{R^{3}}d^{3}\mathbf{x}%
\int_{-\infty}^{+\infty}}dt\Phi(t,\mathbf{x})e^{-\mathbf{i}(\mathbf{k\bullet
x}-\omega t)},\label{x1}\\
\Phi(t,\mathbf{x})  &  ={\frac{1}{(2\pi)^{4}}\int_{R^{3}}d^{3}\mathbf{k}%
\int_{-\infty}^{+\infty}}d\omega\widetilde{\Phi}(\omega,\mathbf{k}%
)e^{\mathbf{i}(\mathbf{k\bullet x}-\omega t)}. \label{x2}%
\end{align}

Inserting Eq.(\ref{x2}) in the HWE we get
\begin{equation}
(\omega^{2}-\mathbf{k}^{2})\widetilde{\Phi}(\omega,\mathbf{k})=0 \label{x3}%
\end{equation}
and we are going to look for solutions of the HWE and Eq.(\ref{x3}) in the
sense of distributions. Then, we rewrite rewrite Eq.(\ref{x3})
\begin{equation}
(\omega^{2}-k_{z}^{2}-\Omega^{2})\widetilde{\Phi}(\omega,\mathbf{k})=0.
\label{x4}%
\end{equation}
where, it is
\begin{align}
\mathbf{\Omega}  &  =k_{x}\mathbf{e}_{x}+k_{y}\mathbf{e}_{y},\nonumber\\
\Omega &  =\left\vert \mathbf{\Omega}\right\vert ^{1/2}=\sqrt{k_{x}^{2}%
+k_{y}^{2}} \label{x4a}%
\end{align}

It is then obvious that any $\widetilde{\Phi}(\omega,\mathbf{k}))$ of the form
\cite{dozi}
\begin{equation}
\widetilde{\Phi}(\omega,\mathbf{k}))=\Xi(\Omega,\beta)\delta\lbrack
\omega-(\beta+\Omega^{2}/4\beta)]\delta\lbrack k_{z}-(\beta-\Omega^{2}%
/4\beta)]\;, \label{x5}%
\end{equation}
where $\Xi(\Omega,\beta)$ is an arbitrary weighting function is a solution of
Eq.(\ref{x4}) since the $\delta$-functions imply that%

\begin{equation}
\omega^{2}-k_{z}^{2}=\Omega^{2}\;. \label{x6}%
\end{equation}

Now, in the dispersion relation given by Eq.(\ref{x6}) define the variables
$\overline{k}$ and $\eta$ by
\begin{equation}
k_{z}=\overline{k}\cos\eta;\quad\cos\eta=k_{z}/\omega,\ \ \omega
>0,\ \ -1<\cos\eta<1\,. \label{x7}%
\end{equation}
and take moreover%
\begin{equation}
\Omega=\overline{k}\sin\eta;\ \ \ \overline{k}>0~~~\text{and }\beta=\frac
{\bar{k}}{2}(1+\cos\eta),~~~0<\eta<\pi/2. \label{x8}%
\end{equation}
Then we have\footnote{The meaning of the subscript $X$ in Eqs.(\ref{x9}) and
(\ref{x10}) will become clear in a while.}%

\begin{equation}
\widetilde{\Phi}_{X}(\omega,\mathbf{k}))=\Xi_{X}(\bar{k},\eta)\delta
(k_{z}-\overline{k}\cos\eta)\delta(\omega-\overline{k}). \label{x9}%
\end{equation}

Now, writing $\mathbf{\rho}=x\mathbf{e}_{x}+y\mathbf{e}_{y}$ and
$\rho=\left\vert \mathbf{\rho}\right\vert ^{1/2}$ we have $\mathbf{\Omega
\bullet\rho}=\Omega\rho\cos\theta$ and using Eq.(\ref{x9}) in Eq.(\ref{x2}) we have

{\small
\begin{equation}
\Phi_{X}(t,\mathbf{x})=\frac{1}{(2\pi)^{4}}\int_{0}^{\infty}d\overline
{k}\;\overline{k}\sin^{2}\eta\left[  \int_{0}^{2\pi}d\theta\,\Xi_{X}%
(\overline{k},\eta)e^{\mathbf{i}\overline{k}\rho\sin\eta\cos\theta}\right]
e^{\mathbf{i}(\overline{k}\cos\eta z-\overline{k}t)}. \label{x10}%
\end{equation}
}

Now choose%

\begin{equation}
\Xi(\overline{k},\eta)=(2\pi)^{3}\frac{z_{0}e^{-\overline{k}z_{0}\sin\eta}%
}{\overline{k}\sin\eta} \label{x11}%
\end{equation}
where $z_{0}>0$ is a constant to get%

\begin{equation}
\Phi_{X}(t,\mathbf{x})=z_{0}\sin\eta\int_{0}^{\infty}d\overline{k}%
e^{-\overline{k}z_{0}\sin\eta}\left[  \frac{1}{2\pi}\int_{0}^{2\pi}d\theta
e^{\mathbf{i}\overline{k}\rho\sin\eta\cos\theta}\right]  e^{\mathbf{i}%
\overline{k}(\cos\eta\,z-t)}\;. \label{x12}%
\end{equation}
Calling $z_{0}\sin\eta=a_{0}>0$, the last equation becomes
\cite{rm1996,rl1997}%

\begin{align}
\Phi_{X}(t,\mathbf{x})  &  =a_{0}\int_{0}^{\infty}d\overline{k}e^{-\overline
{k}a_{0}}J_{0}(\overline{k}\rho\sin\eta)e^{\mathbf{i}\overline{k}(\cos
\eta\,z-t)}\nonumber\\
&  =\frac{a_{0}}{\sqrt{(\rho\sin\eta)^{2}+[a_{0}-\mathbf{i}(z\cos\eta-t)]^{2}%
}} \label{x13}%
\end{align}
which is the now famous $X$-wave solution of the HWE first found ( with a
different approach in \cite{lg}. It is clear that all $\Phi_{X}^{>}$
propagates with speed $c_{1}=1/\cos\eta>1$ in the $z$-direction. This
statement is justified for as can be easily seen is no Lorentz frame where
$\Phi_{XBB_{n}}^{>}$ is at rest. We can easily construct a superluminal
electromagnetic $X$-wave using the Hertz potential method. Indeed, following
\cite{or1998} we write the\ Hertz potential as%
\begin{equation}
\boldsymbol{H}_{X}=\Phi_{X}\boldsymbol{\gamma}^{12}\in\sec%
{\textstyle\bigwedge\nolimits^{2}}
T^{\ast}M\hookrightarrow\sec\mathcal{C\ell(}M,\mathtt{g}\mathcal{)})
\label{x14}%
\end{equation}
The, from Eq.(\ref{9}) we have with $\partial_{\mu\nu}:=\partial^{2}/\partial
x^{\mu}\partial x^{v}$ that
\begin{equation}
\boldsymbol{F}_{X}=\partial_{02}\Phi_{X}\boldsymbol{\gamma}^{01}-\partial
_{01}\Phi_{X}\boldsymbol{\gamma}^{02}+\partial_{32}\Phi_{X}\boldsymbol{\gamma
}^{31}-\partial_{31}\Phi_{X}\boldsymbol{\gamma}^{32}+(\partial_{1}%
^{2}+\partial_{2}^{2})\Phi_{X}\boldsymbol{\gamma}^{12} \label{x15}%
\end{equation}
and since $\boldsymbol{F}^{2}=\boldsymbol{F}\lrcorner\boldsymbol{F+F}%
\wedge\boldsymbol{F}\neq0$ Eq.(\ref{T2}) says that $\mathbf{T}^{0}$ is a
timelike $1$-form field.

\subsection{Pulse Reshaping}

We now want to solve the following Cauchy problem for the Hertz potential
\ with the following initial conditions%

\begin{align}
\boldsymbol{H}_{X}(0,\mathbf{x})  &  =\boldsymbol{\gamma}^{12}\Phi
_{X}=-\mathbf{\sigma}^{12}\frac{a_{0}}{\sqrt{(\rho\sin\eta)^{2}+[a_{0}%
-\mathbf{i}(z\cos\eta)]^{2}}},\nonumber\\
\left.  \frac{\partial}{\partial t}\boldsymbol{H}_{X}(t,\mathbf{x})\right\vert
_{t=0}  &  :=\frac{\partial}{\partial t}\boldsymbol{H}_{X}(0,\mathbf{x}%
)=-\mathbf{\sigma}^{12}\frac{(a_{0}^{2}-\mathbf{i}za_{0}\cos\eta)}{[(\rho
\sin\eta)^{2}+[a_{0}-\mathbf{i}(z\cos\eta)]^{2}]^{3/2}} \label{x17}%
\end{align}
As, well known \cite{vlad} the solution of the proposed Cauchy problem is
given by the following formula%
\begin{equation}
\boldsymbol{H}_{X}(t,\mathbf{x})=-\frac{1}{4\pi}\mathbf{\sigma}^{12}\int
d^{3}\mathbf{x}\left[  G_{R}(t^{\prime},\mathbf{x}^{\prime})\frac{\partial
}{\partial t}\Phi_{X}(t^{\prime},\mathbf{x}^{\prime})-\Phi_{X}(t^{\prime
},\mathbf{x}^{\prime})\frac{\partial}{\partial t}G_{R}(t^{\prime}%
,\mathbf{x}^{\prime})\right]  _{t^{\prime}=0} \label{x18}%
\end{equation}
where
\begin{equation}
G_{R}(t-t^{\prime},\mathbf{x-x}^{\prime})=\frac{\delta(t-t^{\prime}-R)}{R}
\label{X19}%
\end{equation}
$R=\left\vert \mathbf{x-x}^{\prime}\right\vert $ is the retarded Green
function and $R=\left\vert \mathbf{x-x}^{\prime}\right\vert $.

A simple calculation shows that the integral given by Eq.(\ref{1.013}) gives
the \textit{X-}wave solution of the wave equation given by Eq.(\ref{x17}).

This seems remarkable indeed and at first sight a little bit
\textit{paradoxical}. Indeed, the Green function propagates the field value at
any point of the wave in a \textit{causal} way, i.e., with velocity $c=1$. And
yet, e.g., the peak of the \textit{X}-wave moves with superluminal velocity.
This exercise shows a very important fact. The peaks of the wave at two
different instants of time are \textit{not} causally related! The peak at a
given point $\mathbf{x}$ at a given instant of time $t$ is reconstructed by
the points of the wave field which are \ at a distance \ $t-R/c$. This
phenomenon is called \emph{pulse} \textit{reshaping} (or simply reshaping)

What happens if we cutoff \ the Cauchy data \ in Eq.(\ref{x17}) with a window
function like $\Theta(\rho-b)\Theta(|z-\Delta|)$, with $b,\Delta\in\mathbb{R}$
and $\Theta$ the Heaviside function? As a tedious calculation can show, such
truncated $\boldsymbol{F}_{X}$-wave has finite energy. Then, it may be
eventually realized as some real physical phenomenon and indeed experiments
realized by \cite{rm1996} shows that this is the case.\footnote{However, take
notive that the theoretical analysis of \cite{rm1996} are not correct. A
correct analysis is given in \cite{rtx2001,ortx2001}.}

Now, as well known a classical theorem of the theory of hyperbolic
differential equations (see, e.g., \cite{vla} ) warrants that for an initial
field configuration with compact support in space the \emph{front} and the
\emph{rear} of the wave, localized at $t=0$ respectively at $z=\Delta$ and
$z=-\Delta$ will travel with maximum velocity $c=1$\footnote{Before proceeding
it is important to recall that the lateral spreading of a truncated
$\boldsymbol{F}_{X}$-wave happens with a very very small\ velocity (comparing
with the velocity of the front), at least until a distance from the antenna
called the \textit{depth of the field}. For that reason, such waves are
(equivocated) called by some authors non-diffracting waves, which of course is
not the case, because they spread sensibly after travelling a distance greater
than the depth of the field. That such spreading necessarily occurs has been
rigorously proved in \cite{notrod} and it is necessary in order to have energy
conservation in certain well defined situations.}.

But, does reshaping still occurs in this case, thus allowing the peak of the
pulse to travel, at least for a while, with superluminal velocity?

The answer is positive. In order to show that statement we need first to
recall how we can generate a finite aperture approximation to a finite time
pulse $\boldsymbol{H}_{X}(t,\mathbf{x})$. Let as call such object
\[
\boldsymbol{H}_{_{fX}}(t,\mathbf{x})=-\mathbf{\sigma}^{12}\Phi_{X}%
(t,\mathbf{x)\Theta}(\rho-b)\mathbf{\Theta}(\mathbf{|}z\mathbf{-}%
\Delta\mathbf{|)}=-\mathbf{\sigma}^{12}\Phi_{fX}(t,\mathbf{x)}%
\]
. Now, the finite aperture approximation for the Rayleigh-Sommerfeld
of\ diffraction by a plane screen when $\boldsymbol{H}_{X}$ is null in all
space except on the Hertz radiator in the screen located in the $z=0$ plane is
given by%
\begin{equation}
\boldsymbol{H}_{_{fX}}(t,x,y,z)=-\frac{\mathbf{\sigma}^{12}}{2\pi}%
{\textstyle\int\limits_{\mathcal{S}}}
dS^{\prime}\left[  \frac{z}{R^{2}}\frac{\partial}{\partial t^{\prime}}%
\Phi_{fX}(t^{\prime},x^{\prime},y^{\prime},0)+\frac{1}{R}\Phi_{_{fX}%
}(t^{\prime},x^{\prime},y^{\prime},0)\right]  _{t^{\prime}=t-R} \label{X18}%
\end{equation}

We see that the peak of $\boldsymbol{H}_{fX}$ at say $z=L$ is produced from
rings at the $z=0$ at different at different retarded times which can be read
from Eq.(\ref{X18}) and for a finite time interval $T=2\Delta$.

One \textit{qualitative} prediction that can be immediately given in such a
case is that the velocity of the peak must slowly as it propagates along the
$z$-axis, since it will catch the front which is moving with velocity $c=1$.

\begin{remark}
The experiment described in \emph{\cite{rl1997}} \emph{(}and also in
\emph{\cite{or1998})} did not look at that effect, since it supposed that the
Hertz radiator was on from $-\infty<t<\infty.$ However, in that paper it was
presented a \textit{X}-wave solution for Maxwell equations called the
SEXW\textit{ }\emph{(}superluminal electromagnetic \textit{X}-wave\emph{)} in
\emph{\cite{rl1997}} , and simulations for the motion of finite aperture
approximations for that waves \emph{(}\textit{FAASEXW}\emph{)} have been
investigated showing that their peaks can propagate \emph{(}with well designed
antennas\emph{)} for long distances with superluminal velocities.
\end{remark}

\begin{remark}
Waves of this type \emph{(}using antennas different from the simple one
detailed above\emph{)} have been produced originally in both the optical
\emph{\cite{saari}}, as well the microwave regions \emph{\cite{mr}}. In that
last paper the phenomenon of the slowing of the velocity of the peak was
observed. Since the authors of \emph{\cite{mr}} was unaware of the reshaping
phenomenon occurring \emph{even} in vacuum they mislead their readers,
claiming to have detected a \textit{genuine} superluminal
motion\emph{\footnote{What has been measured in the experiment \cite{mr} was
not the velocity of the \textit{front} of the wave, which must be $c=1$
according to the the correct theory of the phenomenon. The reason for thir
claim has to do with the detection threshold of their measurement device. In
this respect see \cite{rtx2001,ortx2001}. where some details of the experiment
and its theoretical analysis are presented.}}, what has not been the case. We
emphasize, what they measured \ was nothing more than the velocity of the
\textit{peaks} of \textit{FAASEXWs} \ while moving along the $z$-axis. The
peaks at two different positions (at different times) were not causally
connected, something that is obviously if we feed the second member of
\emph{Eq.(\cite{X8})} with initial conditions%
\begin{align}
\boldsymbol{H}_{fX}(0,\mathbf{x})  &  =-\mathbf{\sigma}^{12}\Phi
_{X}=-\mathbf{\sigma}^{12}\frac{a_{0}\Theta(\rho-b)\Theta(|z-\Delta|)}%
{\sqrt{(\rho\sin\eta)^{2}+[a_{0}-\mathbf{i}(z\cos\eta)]^{2}}},\nonumber\\
\left.  \frac{\partial}{\partial t}\boldsymbol{H}_{fX}(t,\mathbf{x}%
)\right\vert _{t=0}  &  :=\frac{\partial}{\partial t}\boldsymbol{H}%
_{fX}(0,\mathbf{x})=-\mathbf{\sigma}^{12}\frac{(a_{0}^{2}-\mathbf{i}za_{0}%
\cos\eta)\Theta(\rho-b)\Theta(|z-\Delta|)}{[(\rho\sin\eta)^{2}+[a_{0}%
-\mathbf{i}(z\cos\eta)]^{2}]^{3/2}} \label{x19a}%
\end{align}
in order to get $\boldsymbol{H}_{\boldsymbol{fX}}$ using \emph{Eq.(\ref{x18})}

So, what has been measured in the experiment \emph{\cite{mr}} was not the
velocity of the \textit{front} of the wave, which must be $c=1$ according to
the the correct theory of the phenomenon. The reason for their wrong claim has
to do with the detection threshold of their measurement device. In this
respect see \emph{\cite{rtx2001,ortx2001}} where some details of the
experiment and its theoretical analysis are presented.
\end{remark}

\begin{remark}
It is important to emphasize here that the reshaping phenomena that as we just
proved may happens even in vacuum is crucial to explain the existence of
superluminal \emph{(}and even negative) group velocities\ for electromagnetic
waves propagating in a dispersive medium and wave guides\emph{\footnote{Like
e.g, in the famous Nintz experiment \cite{nimtz} where he claimed to have sent
an electromagntic field configration through a wave guide with speed greater
than the light speed in vacuum. Of course, a careful theoretical analysis of
his experiment predicts that he did not measurethe velocity of the front of
his generated wave. Due to the detection threshold of his device he can only
detect the arrival of the peak of his signal which due to the reshapping
phenomenon travels at superluminal speed.}}. We refer the reader to the
following original papers\footnote{See also \cite{hzr} for a more recent
account of experiments and some theoretical ideas (not all in agreement with
our studies on this issue).}
\emph{\cite{c1,ckk,dl,diener1,diener2,diener3,emig,gcumber,l,michiao,mende,steinberg}
and} for the description of how superluminal \emph{(}and even negative\emph{)
}group velocities appear\footnote{For the case of the tunneling of electrons
through a barrier an explanation of the superluminal velocity of the peak of
the electron wave function is again the reshaping phenomenon that seems to be
universal. A careful analysis is given in \cite{dlg}. These authors show that
in this case the integral lines of the Dirac current seems to correspond
indeed to the possible electron trajectories.}, how this is explained by the
reshaping phenomenon and how it is possible to prove that even with
superluminal velocities the energy in a dispersive medium or wave guide always
travel with speed $v_{\varepsilon}\leq1$.
\end{remark}

\subsection{Character of $\mathbf{T}_{0}$ for $\boldsymbol{F}_{_{fX}%
}(t,\mathbf{x})$. How Fast Travels the Energy of $\boldsymbol{F}_{_{fX}%
}(t,\mathbf{x})$?}

From Eq.(\ref{x5}) we can easily construct $\boldsymbol{F}_{fX}(t,\mathbf{x})$
and then verify that%

\begin{equation}
\boldsymbol{F}_{\boldsymbol{fX}}^{2}=\boldsymbol{F}_{\boldsymbol{fX}}%
\lrcorner\boldsymbol{F}_{fX}\text{ }\boldsymbol{+}\text{ }\boldsymbol{F}%
_{fX}\wedge\boldsymbol{F}_{fX}\neq0 \label{x20}%
\end{equation}
implying that energy-momentum $1$-form $\mathbf{T}_{0}=-1/2\boldsymbol{F}%
_{fX}\boldsymbol{\boldsymbol{\gamma}^{0}F}_{fX}$ is timelike (as recalled in
Section 3). But, of course since the front and the rear of $\boldsymbol{F}%
_{_{fX}}(t,\mathbf{x})$ moves with $c=1$ it is not reasonable to suppose that
the energy flows at superluminal speed. Let us them analyze our problem with
care and see if we can get a possible solution. To start recall that
$\mathbf{T}_{0}$ satisfies the fundamental differential conservation equation%
\begin{equation}
\delta\mathbf{T}_{0}=0 \label{x20a}%
\end{equation}
Thus, if we define%
\begin{equation}
\mathbf{T}_{0}^{\prime}=\mathbf{T}_{0}+\delta\mathcal{X} \label{x20b}%
\end{equation}
where $\mathcal{X}$ is a $0$-form field then we still have that $\delta
\mathbf{T}_{0}^{\prime}=0;$

Thus as well know the energy velocity is supposed to be given by
\begin{equation}
v_{\varepsilon}=\frac{\left\vert \mathbf{P}\right\vert }{u} \label{x21}%
\end{equation}
where $u$ and $\mathbf{P}$\textbf{ }are giving by Eqs.(\ref{a19}),(\ref{a20})
and (\ref{a21}) calculated with $\mathbf{T}_{0}$. It follows from the fact
that $\mathbf{T}_{0}$ is timelike $v_{\varepsilon}<1$. Recall now that it is
the integral form of the energy-momentum conservation law is%

\begin{equation}
\frac{\partial}{\partial t}\left\{  \int_{\mathcal{V}}d^{3}\mathbf{x}\frac
{1}{2}(\mathbf{E}^{2}+\mathbf{B}^{2})\right\}  =\oint_{\mathcal{S}}%
d\mathbf{S}\bullet\mathbf{P} \label{x22}%
\end{equation}
that really describes the way in which energy flows. In Eq.(\cite{x22}) it is
clear that we can make a choice of $\mathcal{X}$ in Eq.(\ref{x20}) such that
the energy density continues to be given by $\frac{1}{2}(\mathbf{E}%
^{2}+\mathbf{B}^{2})$ and such that $\mathbf{P\mapsto P+P}^{\prime}$ with
$\nabla\bullet\mathbf{P}^{\prime}=0$ then there is no alteration of the flow
of the energy contained in $\mathcal{V}$. We can verify using Eqs.(\ref{a19}%
),(\ref{a20}) and (\ref{a21}) \ that we get such a situation if we choose a
$\mathcal{X}$ satisfying $\partial_{0}\mathcal{X}=0$ and $\nabla
^{2}\mathcal{X}=0$. Indeed, in this case we have the $u$ as in Eq.(\cite{x21})
and $\mathbf{P}^{\prime}=\mathbf{P+\partial}^{i}\mathcal{X}\mathbf{\sigma}%
_{i}.$ Thus with a thoughtful solution of $\nabla^{2}\mathcal{X}=0$ we can get%

\begin{equation}
v_{\varepsilon}^{\prime}=\frac{|\mathbf{P+P}^{\prime}|}{u}=1. \label{x23}%
\end{equation}

\begin{remark}
We observe that the $\boldsymbol{F}_{fX}$ that we exhibit above is a classical
electromagnetic field configuration, which if described in terms of photons
\cite{mw} has an undetermined number of them, A such, of course we cannot even
in principle to think that for this case the integral lines of $\mathtt{g}%
(\mathbf{T}_{0},~)$ describe any photon trajectory;
\end{remark}

\section{Character of $\mathbf{T}^{0}$ for a Static Electric Plus Magnetic
Field Configuration}

Consider a spherical conductor in electrostatic equilibrium with uniform
superficial charge density (total charge $Q$) and with a dipole magnetic
moment of radius $R$. Then, we have (with $\mathbf{e}_{\mathbf{r}}%
:=\mathbf{x}/\left\vert \mathbf{x}\right\vert $) for $r=\left\vert
\mathbf{x}\right\vert >R$%

\begin{equation}
\mathbf{E}=\frac{Q}{r^{2}}\mathbf{e}_{\mathbf{r}}\ \ ;\ \ \mathbf{B}=\frac
{C}{r^{3}}(2\cos\theta\,\mathbf{e}_{\mathbf{r}}+\sin\theta
\,\mbox{\boldmath $\theta$}) \label{19}%
\end{equation}
which of course, for $r>R$ satisfy the free Maxwell equations. We have%

\begin{equation}
\mathbf{P}=\mathbf{E}\times\mathbf{B}=\frac{CQ}{r^{6}}\sin\theta
\,\mbox{\boldmath
$\varphi$}\ \ ,\ \ u=\frac{1}{2}(Q^{2}+C^{2}(3\cos^{2}\theta+1)). \label{20}%
\end{equation}
Thus%

\begin{equation}
\frac{|\mathbf{P}|}{u}=\frac{2CQ\sin\theta}{Q^{2}+C^{2}(3\cos^{2}\theta
+1)}\leq1 \label{21}%
\end{equation}
and since the fields are static the conservation law Eq.(\ref{x22}) continues
to hold true, and the first member is obviously null and thus for any closed
surface containing the spherical conductor we have
\begin{equation}
\oint_{\mathcal{S}}d\mathbf{S}\bullet\boldsymbol{P}=0. \label{22}%
\end{equation}

So, we arrive at the conclusion that in static electric plus magnetic field
something carrying energy and momentum is in motion around the charged
magnetized sphere. If this is true then there must be an angular momentum
stocked in the field. And in fact this angular momentum vector is
\cite{rvr1993}:%
\begin{equation}
\int d\mathbf{v(x\times}\boldsymbol{P})=\frac{1}{3}\frac{CQ}{4\pi R^{2}%
}\mathbf{e}_{\mathbf{z}}. \label{23}%
\end{equation}

It has been shown in \cite{sharma} that if the sphere is slowly discharged (in
order to not radiate) through its south pole it acquires mechanical angular
momentum exactly equal to the electromagnetic angular momentum\footnote{A more
general and important case showing how to get energy from an external magnetic
field in order to put a motor in motion has been discussed in \cite{rc2014}%
}.The fact that static electric plus magnetic fields stocked electromagnetic
angular momentum has been verified experimentally by Lahoz and Graham in 1989
\cite{lgram}.

With the results of this and the previous section we think it is time to quote
Stratton \cite{stratton} that in section 2.19 of his classical book said:

\begin{quotation}
By this standard there is every reason to retain the Poynting-Heaviside
viewpoint until a clash with new experimental evidence shall call for its revision.
\end{quotation}

Well, here we present two cases where a revision is necessary. But with this
revision the idea of associating photon trajectories to the integral lines of
$\mathtt{g}(\mathbf{T}_{0},~)$ becomes of course meaningless. It is necessary
to check if association of photon trajectories as in the Bohm like theory
presented in \cite{fh2016} can predict correct the photon trajectories for the
electromagnetic field configuration $\boldsymbol{F}_{\boldsymbol{fX}}$. We
also suggest here that an experiment should be done to clarify which kind of
trajectories can be determined in this case using, e.g., the techniques of the
Toronto experiment \cite{mrf}.

\section{How $\mathrm{i}=\sqrt{-1}$ Enters Maxwell Theory}

\subsection{Helicity States}

Starting from ME the most simple solution are PWS. Choose a solution, moving,
e.g., in the $\mathbf{z}$-direction, given by%

\begin{equation}
\boldsymbol{F}=\mathbf{f}e^{\boldsymbol{\gamma}^{5}k\cdot x} \label{2.39}%
\end{equation}
with $k=k_{\mu}\boldsymbol{\gamma}^{\mu}$, $k_{1}=k_{2}=0$, $x=x^{\mu
}\boldsymbol{\gamma}_{\mu}$, and $\mathcal{\boldsymbol{F}}$ a constant 2-form.
It is immediately to conclude that $k\boldsymbol{F}=0$ and $k\cdot k=0$ and then%

\begin{equation}
k\cdot k=0\ \Rightarrow\ k_{0}=\pm|\mathbf{k}|. \label{2.43}%
\end{equation}

It is interesting to understand the fundamental role of the volume element
$\boldsymbol{\gamma}^{5}$ (duality operator) in electromagnetic theory. In
particular since $e^{\boldsymbol{\gamma}^{5}k\cdot x}=\cos k\cdot
x-\boldsymbol{\gamma}_{5}\sin k\cdot x$, we see that%

\begin{equation}
\boldsymbol{F}=\mathbf{f}\cos kx-\boldsymbol{\gamma}_{5}\mathbf{f}\sin kx.
\label{2.45}%
\end{equation}
Writing $\boldsymbol{F}=\mathbf{E}+\mathbf{iB}$, with%
\begin{gather}
\mathbf{i}\equiv\boldsymbol{\gamma}_{5}=\mathbf{e}_{1}\mathbf{e}_{2}%
\mathbf{e}_{3},\nonumber\\
\mathbf{e}_{i}=\boldsymbol{\gamma}_{i}\boldsymbol{\gamma}_{0} \label{2.45a}%
\end{gather}
and choosing, e.g., $\mathbf{f}=\mathbf{e}_{1}+\mathbf{ie}_{2}$,
$\mathbf{e}_{1}\bullet\mathbf{e}_{2}=0$, $\mathbf{e}_{1}$, $\mathbf{e}_{2}$
constant $2$-forms to be treated in calculations below (with the appropriate
symbols) as $1$-forms in the Pauli subbundle $\mathcal{C\ell}^{0}%
(M,\mathtt{g})$ of $\mathcal{C\ell}(M,\mathtt{g})$%

\begin{equation}
(\mathbf{e}+\mathbf{ib})=\mathbf{e}_{1}\cos k\cdot x+\mathbf{e}_{2}\sin
kx+\mathbf{i}(-\mathbf{e}_{1}\sin k\cdot x+\mathbf{e}_{2}\cos k\cdot x).
\label{2.46}%
\end{equation}
This equation is important because it shows that we must take care with the
$i=\sqrt{-1}$ that appears in usual formulations of Maxwell Theory using
complex electric and magnetic fields. The $i=\sqrt{-1}$ in many cases unfolds
a secret that can only be known through Eq.(\ref{2.46}). From the fact that
$k\boldsymbol{F}=0$ we can also easily show (see below) that $\mathbf{k}%
\bullet\mathbf{E}=\mathbf{k}\bullet\mathbf{B}=0$, \textit{i.e.\/}, \emph{PWS}
of \emph{ME} are \emph{transverse waves}. From ME satisfied by $\boldsymbol{F}%
$ we have immediately that
\[
k_{0}\boldsymbol{F}=\mathbf{k}\boldsymbol{F}.
\]

On the other hand we can write $k\boldsymbol{F}=0$ as%

\begin{equation}
k\boldsymbol{\gamma}_{0}\boldsymbol{\gamma}_{0}\boldsymbol{F\gamma}_{0}=0
\label{2.47}%
\end{equation}
and since $k\boldsymbol{\gamma}_{0}=k_{0}+\mathbf{k}$,
defining$\ \boldsymbol{F}^{\ast}=\boldsymbol{\gamma}_{0}\boldsymbol{F\gamma
}_{0}=-\mathbf{E}+\mathbf{iB}$ we have
\begin{equation}
k_{0}\boldsymbol{F}^{\ast}=-\mathbf{k}\boldsymbol{F}^{\ast}. \label{2.48}%
\end{equation}
So, we see that $^{\ast}$ plays the role of the operator of space conjugation
\cite{hestenes}. Of course, for $\mathbf{f}=\mathbf{e}+\mathbf{ib}$ we have%

\begin{equation}
\mathbf{f}^{\ast}=-\mathbf{e}+\mathbf{ib}\ \ ;\ \ k_{0}^{\ast}=k_{0}%
\ \ ;\ \mathbf{k}^{\ast}=-\mathbf{k}. \label{2.49}%
\end{equation}
We can now interpret the two solutions of $k^{2}=0$, \textit{i.e.\/},
$k_{0}=|\mathbf{k}|$ and $k_{0}=-|\mathbf{k}|$ as corresponding to the
solutions
\begin{equation}
\text{(a) }k_{0}\mathbf{f}=\mathbf{kf},~~~~\text{(b) }k_{0}\mathbf{f}^{\ast
}=-\mathbf{kf}^{\ast} \label{2.49x}%
\end{equation}
where $\mathbf{f}$ and $\mathbf{f}^{\ast}$ correspond in quantum theory to
\textquotedblleft photons\textquotedblright\ which are of positive or negative
\emph{helicities}. This will be important for our discussion in Section 4.

Observe that from Eq.(\ref{2.49x}(a)) we can write%
\begin{equation}
k_{0}\mathbf{e+}k_{0}\mathbf{ib=k}(\mathbf{e}+\mathbf{ib})=\mathbf{k}%
\lrcorner\mathbf{e+k}\wedge\mathbf{e+k\lrcorner(ib})+\mathbf{k}\wedge
\mathbf{(ib}) \label{2.49xx}%
\end{equation}
and comparing the grades in both members, taking into account that
$\mathbf{k}\wedge\mathbf{(ib})=\mathbf{i}$ $(\mathbf{k}\lrcorner\mathbf{b})$
and $\mathbf{k}\lrcorner\mathbf{(ib})=\mathbf{i(\mathbf{k}}\wedge
\mathbf{\mathbf{b})}$ we get that
\begin{equation}
\mathbf{k}\lrcorner\mathbf{e=k}\lrcorner\mathbf{b}=0 \label{2.49xxx}%
\end{equation}
and .%
\begin{equation}
k_{0}(\mathbf{e+ib)=k\wedge e+i(k\wedge b}) \label{2.49b}%
\end{equation}
and interpreting the objects in Eq.(\ref{2.49b}) as objects with values in the
Pauli algebra it is $\mathbf{k\times e:=-i(k\wedge e)}$ and we can write
\begin{equation}
-\mathbf{i}k_{0}\mathbf{f}=\mathbf{(k\times e)+i(k\times b})=\mathbf{k\times
f.} \label{2.49a}%
\end{equation}

\begin{remark}
In what follows we will \ impose that
\begin{equation}
\left\vert \mathbf{f}\right\vert ^{2}=\mathbf{f\bullet f}^{\ast}=1
\label{2.499}%
\end{equation}
and write
\begin{equation}
\mathbf{f}_{+}:=\mathbf{f}\text{,~~~}\mathbf{f}_{-}:=\mathbf{f}^{\ast}
\label{2.49v}%
\end{equation}
corresponding respectively to right and left hand photons.
\end{remark}

\subsection{How a Duality Rotations Turns Up in a Spatial Rotation}

The plane wave solutions (PWS) of ME are usually obtained by looking for
solutions to these equations such that the potential $\boldsymbol{A}=A_{\mu
}\boldsymbol{\gamma}^{\mu}$ is in the \textit{Coulomb} (or \textit{radiation})
gauge, i.e.,
\begin{equation}
A_{0}=0,\text{ }\partial_{i}A^{i}:=\nabla\bullet\mathbf{A}=0 \label{10.4.1n}%
\end{equation}

Eq.(\ref{10.4.1n}), of course, implies that $\boldsymbol{\partial}%
\cdot\boldsymbol{A}=-\delta\boldsymbol{A}=0$, i.e., the Lorenz gauge condition
is also satisfied. Now, in the absence of sources, $\boldsymbol{A}$ and
$\boldsymbol{F}=\boldsymbol{\partial}\wedge\boldsymbol{A}=d\boldsymbol{A}$
satisfy, respectively the homogeneous wave equation (HWE) and the free ME
\begin{align}
\square\boldsymbol{A}  &  =0,\label{10.4.2n}\\
\boldsymbol{\partial F}  &  =0
\end{align}

The PWS can be obtained directly from the free ME ($\boldsymbol{\partial F}%
=0$) once we suppose that $\boldsymbol{F}$ is a null field, i.e.,
$\boldsymbol{F}^{2}=0$. However, for the purposes we have in mind we think
more interesting to find these solutions by solving $\square\boldsymbol{A}=0$
with the subsidiary condition given by Eq.(\ref{10.4.1n}).

In order to do that we introduce besides $\{x^{\mu}\}$ another set of
coordinate $\{x^{\prime\mu}\}$ also in the Einstein-Lorentz-Poincar\'{e} gauge
which are also a $(n\boldsymbol{A}cs|\partial/\partial x^{0})$, such that ,
$x^{\prime0}=x^{0}$, $x^{\prime i}=R_{j}^{i}x^{j}$. Putting
$\boldsymbol{\varepsilon}^{\mu}=dx^{\prime\mu}$ we then have
\begin{equation}
\boldsymbol{\varepsilon}^{\mu}=R\boldsymbol{\gamma}^{\mu}\tilde{R};\text{
}\varepsilon^{0}=\boldsymbol{\gamma}^{0}\text{, }\varepsilon^{i}%
\neq\boldsymbol{\gamma}^{i}\text{,}%
\end{equation}
where the (constant) $R\in\sec\mathrm{S}$\textrm{$\boldsymbol{P}$}%
$\mathrm{in}_{3}(M)$ generates a global rotation of the space axes. We also
write
\begin{gather}
\boldsymbol{\varepsilon}_{i}\boldsymbol{\varepsilon}_{j}\boldsymbol{\equiv
\varepsilon}_{ij}\boldsymbol{,\ \ \varepsilon}_{i}\boldsymbol{\varepsilon}%
_{0}=\mathbf{e}_{i},,\text{ \ \ }\mathbf{e}_{i}\mathbf{e}_{j}\equiv
\mathbf{e}_{ij}=-\boldsymbol{\varepsilon}_{ij},\text{ \ \ }\mathbf{i}%
=\mathbf{e}_{1}\mathbf{e}_{2}\mathbf{e}_{3},\nonumber\\
i,j=1,2,3\text{ and }i\neq j.
\end{gather}

We consider next the following two linearly independent monochromatic plane
solutions $\boldsymbol{A}^{(i)}$, $i=1,2$, of Eq.(\ref{10.4.2n}) satisfying
the subsidiary condition giving by Eq.(\ref{10.4.1n}) and moving in the
$\mathbf{e}_{3}$ direction,
\begin{gather}
\boldsymbol{A}^{(i)}=\exp\left[  \frac{(-1)^{i+1}\boldsymbol{\varepsilon}%
_{12}\bar{\phi}_{i}}{2}\right]  \varepsilon_{1}\exp\left[  \frac
{-(-1)^{i+1}\boldsymbol{\varepsilon}_{12}\bar{\phi}_{i}}{2}\right]
=\exp\left[  \boldsymbol{\varepsilon}_{12}(-1)^{i+1}\bar{\phi}_{i}\right]
\boldsymbol{\varepsilon}_{1},\nonumber\\
\bar{\phi}_{i}:M\rightarrow\mathbb{R}\text{, }\mathfrak{e}\mapsto\bar{\phi
}_{i}(x)=k_{\mu}^{\prime}x^{\prime\mu}+\bar{\varphi}_{i}=\omega T-\mathbf{k}%
^{\prime}\cdot\mathbf{x}^{\prime}+\bar{\varphi}_{i}\text{, }\omega
=|\mathbf{k}^{\prime}|,\text{ }\nonumber\\
\mathbf{k}^{\prime}=\omega\mathbf{e}_{3}.
\end{gather}
where the $\bar{\varphi}_{i}$ are real constants, called the initial phase.
Since A$_{0}^{(i)}=0$ we write,%
\begin{equation}
\mathbf{A}^{(i)}=\boldsymbol{A}^{(i)}\varepsilon_{0}=\exp\left[
-(-1)^{i+1}\mathbf{e}_{2}\mathbf{e}_{1}\bar{\phi}_{i}\right]  \mathbf{e}%
_{1}=\mathbf{e}_{1}\exp\left[  (-1)^{i+1}\mathbf{e}_{2}\mathbf{e}_{1}\bar
{\phi}_{i}\right]  ,
\end{equation}

Now,
\begin{align}
\boldsymbol{F}^{(i)}  &  =\partial\wedge\boldsymbol{A}^{(i)}=\partial
\wedge\boldsymbol{A}^{(i)}=\mathbf{\partial}\boldsymbol{\varepsilon}%
_{0}\boldsymbol{\varepsilon}_{0}\boldsymbol{A}^{(i)}\nonumber\\
&  =(\partial_{t}-\nabla)(-\boldsymbol{A}^{(i)}). \label{10.4.8n}%
\end{align}

We calculate in details $\boldsymbol{F}^{(1)}$ in order for the reader to see
explicitly how $\mathbf{i}=\mathbf{e}_{1}\mathbf{e}_{2}\mathbf{e}_{3}$ enters
in the classical formulation of the electromagnetic field. We have,
\begin{align}
\boldsymbol{F}^{(1)}  &  =-\omega\left[  \mathbf{e}_{1}\mathbf{e}%
_{2}\mathbf{e}_{1}\exp(e_{21}\bar{\phi}_{1})-\mathbf{e}_{3}\mathbf{e}%
_{1}\mathbf{e}_{2}\mathbf{e}_{1}\exp(\mathbf{e}_{21}\bar{\phi}_{1})\right]
\nonumber\\
&  =\omega\left[  \mathbf{e}_{2}-\mathbf{ie}_{1}\right]  \exp(-\mathbf{e}%
_{2}\mathbf{e}_{1}\bar{\phi}_{1})\nonumber\\
&  =\omega\left[  \mathbf{e}_{1}+\mathbf{ie}_{2}\right]  \exp(-\mathbf{i}%
\phi_{1})\nonumber\\
\phi_{1}  &  =\omega t-\mathbf{k}^{\prime}\cdot\mathbf{x}^{\prime}+\varphi
_{1},\varphi_{1}=\bar{\varphi}_{1}-\frac{\pi}{2} \label{fi}%
\end{align}

This formula shows how a \emph{duality rotation} originally appearing in the
formula for $\boldsymbol{F}$ turns up into \emph{spatial rotation} in that
formula a really non trivial result. Besides that \ Besides that Eq.(\ref{fi})
shows that representation of electromagnetic fields through complex fields is
a simple representation of what has been just found.

\subsection{Schr\"{o}dinger Form of Maxwell Equation}

Here we derive a three dimensional representation of the free ME, first
presented Riemann\footnote{See details in \cite{weber}} and rediscovered by
Silberstein \cite{sil1,sil2},Bateman \cite{bat}, Majorana\footnote{See the
article \cite{mrb}.} but obtained in a completely different way from the one
given below. Our starting point is Maxwell equation written in the Clifford
bundle, i.e., \ $\boldsymbol{\partial F}=0$ which since it is also satisfied
by $\boldsymbol{\gamma}_{5}\boldsymbol{F}$ can be rewritten (in the case
$J=0$) in the following equivalent ways in the even subbundle $\mathcal{C}%
\ell^{0}(M,\mathtt{\eta})$ of $\mathcal{C}\ell(M,\mathtt{\eta})$ once we write
$\mathbf{\sigma}^{i}=\boldsymbol{\gamma}^{i}\boldsymbol{\gamma}^{0}$ and
$\mathbf{\sigma}^{0}=1$
\begin{align}
\boldsymbol{\partial}(\boldsymbol{\gamma}_{5}\boldsymbol{F})  &
=0,\nonumber\\
\mathbf{i\sigma}^{\mu}\partial_{\mu}\boldsymbol{F}  &  =0,\label{10.2.9}\\
(\mathbf{i}\frac{\mathbf{\sigma}^{0}}{2})\frac{\partial}{\partial
T}\boldsymbol{F}  &  =-\mathbf{i}\frac{\mathbf{\sigma}^{i}}{2}\partial
_{i}\boldsymbol{F}.\nonumber
\end{align}

Now, we recall that $\left[  \mathbf{\sigma}^{i}/2,\mathbf{\sigma}%
^{j}/2\right]  =\mathbf{i}\varepsilon_{k}^{ij}\mathbf{\sigma}^{k\text{ }}/2$,
i.e., the set \{$\mathbf{\sigma}^{i}/2$\} is a basis for any $\mathfrak{e}\in
M$ of the Lie algebra $\mathrm{su}(2)$ of $\mathrm{SU}(2)$, the universal
covering group of \textrm{SO}$_{3}$, the special rotation group in three
dimensions. A three dimensional representation of $\mathrm{su}(2)$ is given by
the Hermitian matrices
\begin{equation}
\mathbf{\hat{\Sigma}}^{\boldsymbol{p}}=\left[
\begin{array}
[c]{lll}%
0 & -\mathrm{i}\delta^{\boldsymbol{p}3} & \mathrm{i}\delta^{\boldsymbol{p}2}\\
\mathrm{i}\delta^{\boldsymbol{p}3} & 0 & -\mathrm{i}\delta^{p1}\\
-\mathrm{i}\delta^{\boldsymbol{p}2} & \mathrm{i}\delta^{\boldsymbol{p}1} & 0
\end{array}
\right]  \label{10.2.10}%
\end{equation}
and
\begin{equation}
\left[  \mathbf{\hat{\Sigma}}^{\boldsymbol{p}},\mathbf{\hat{\Sigma}}%
^{q}\right]  =\mathrm{i}\epsilon_{\cdot\cdot r}^{\boldsymbol{p}q\cdot
}\mathbf{\hat{\Sigma}}^{r}. \label{10.2.10.0}%
\end{equation}

Writing moreover $\mathbf{\hat{\Sigma}}^{0}=\mathbf{I}_{3}$ for the three
dimensional unitary matrix and defining
\begin{equation}
\mathbf{\boldsymbol{F}}=\left[
\begin{array}
[c]{l}%
E_{1}+\mathrm{i}B_{1}\\
E_{2}+\mathrm{i}B_{2}\\
E_{3}+\mathrm{i}B_{3}%
\end{array}
\right]  , \label{10.2.11}%
\end{equation}
we can obtain ME in three dimensional form with the substitutions
\begin{equation}
\frac{1}{2}\mathbf{\sigma}^{\mu}\mapsto\mathbf{\hat{\Sigma}}^{\mu
},~~~\mathbf{i}\mapsto\mathrm{i}=\sqrt{-1},~~~\boldsymbol{F}\mapsto
\mathbf{\boldsymbol{F}}. \label{10.2.12}%
\end{equation}
in Eq.(\ref{10.2.9}). We get
\begin{equation}
\mathrm{i}\frac{\partial}{\partial T}\mathbf{\boldsymbol{F}}=-\mathrm{i}%
\mathbf{\hat{\Sigma}}\bullet\mathbf{\nabla\boldsymbol{F}} \label{10.2.12.1}%
\end{equation}
Note the \textit{doubling} of the representative of the unity element of
$\mathcal{C}\ell^{0}(M,\mathtt{\eta})$ when going to the three dimensional
representation. This corresponds to the fact that in relativistic quantum
field theory, the $2\times2$ matrix representation of Eq.(\ref{10.2.9})
(projected in the idempotent $\frac{1}{2}(1+\mathbf{\sigma}^{3})$)
represents\footnote{Of course, it is necessary for the quantum mechanical
interpretation to multiply both sides of eq.(\ref{10.2.12.1}) by $\hbar$, the
Planck constant.} the wave equation for a single quantum of a massless spin
1/2 field, whereas Eq.(\ref{10.2.12.1}) represents the wave equation for a
single quantum of a massless spin 1 field\footnote{Indeed in quantum mechanics
the Pauli matrices $\mathbf{\sigma}_{i}$ and the matrices $\mathbf{\Sigma}%
_{i}$ are the quantum mechanical spin operators and
\[
\sum_{i=1}^{3}(\mathbf{\hat{\sigma}}_{i})^{2}=\frac{1}{2}(1+\frac{1}{2}%
)=\frac{3}{4},\text{ }\sum_{i=1}^{3}(\mathbf{\hat{\Sigma}}_{i})^{2}%
=1.(1+1)=2.
\]
}.

\begin{remark}
\emph{Eq.(\ref{10.2.12.1}) }has been used in several papers that discusses the
possibility of writing photon wave functions \emph{\cite{bia} and the question
of the maximum localizability of photons, e.g., in \cite{as,jajan,sch}. We are
not going to discuss these papers here, for instead we shall present an
alternative path to the photon wave function based on a paper by
Bialynicki-Birula \cite{biab}, but using Maxwell equation in its form given in
the Clifford bundle }$\mathcal{C\ell(}M,\mathtt{g})$\emph{ by Eq.(\ref{4n}) or
its form as given in the Clifford bundle }$\mathcal{C\ell}^{0}\mathcal{(}%
M,\mathtt{g})$ by \emph{Eq.(\ref{10.2.9}).}
\end{remark}

\section{Enter the Quantum Schr\"{o}dinger Equation for the Photon}

Once we have discovered that there are two different polarizations associated
to \ a PWS of ME we can write (once we fix an inertial reference frame and use
coordinates in ELP gauge) a general solution of Eq.(\ref{4n}) as%
\begin{equation}
\boldsymbol{F}(t,\mathbf{x})=\frac{1}{(2\pi)^{4}}\int d^{4}k\delta
(k^{2})\mathbf{f}(k)U(k)e^{-\gamma^{5}k\cdot x}. \label{pn0}%
\end{equation}
We choose the $2$-forms $\mathbf{f}$ such that $\mathbf{f}(k_{0},\pm
\mathbf{k})=\mathbf{f}_{\pm}\mathbf{(k)}$ describe the two possible
polarizations discussed above. Thus, we can write
\begin{gather}
\boldsymbol{F}(t,\mathbf{x})=\boldsymbol{F}_{+}(t,\mathbf{x})+\boldsymbol{F}%
_{-}(t,\mathbf{x})\nonumber\\
=\frac{1}{(2\pi)^{3}}\int d^{3}\mathbf{k}\frac{1}{2\omega\mathbf{_{\mathbf{k}%
}}}\{\mathbf{f}_{+}\mathbf{(k)}u_{+}\left(  \mathbf{k}\right)  +\frac
{1}{2\omega\mathbf{_{\mathbf{k}}}}\mathbf{f}_{-}\mathbf{(k)}u_{-}\left(
\mathbf{k}\right)  \}e^{-\mathbf{i}\omega_{\mathbf{k}}t+\mathbf{k\bullet x}}]
\label{pn1}%
\end{gather}
with $\omega_{\mathbf{k}}=k_{0}=\left\vert \mathbf{k}\right\vert $ and where
$u_{\pm}:\mathbb{R}^{3}\rightarrow\mathbb{R}\oplus\mathbf{i}\mathbb{R}$ are
\textquotedblleft complex\textquotedblright\ functions conveniently chosen and
whose nature investigate in what follows. In order to do that we recall once
again that the classical energy density is given by%
\begin{equation}
\overset{c}{T}_{00}(t,\mathbf{x)}=\boldsymbol{F}(t,\mathbf{x})^{\ast}%
\bullet\boldsymbol{F}(t,\mathbf{x})=\left\vert \boldsymbol{F}(t,\mathbf{x}%
)\right\vert ^{2}=\frac{1}{2}\left(  \mathbf{E}^{2}+\mathbf{B}^{2}\right)
\label{pn2}%
\end{equation}
and as we already saw, for a monochromatic PWS its energy $%
{\textstyle\int}
d^{3}\mathbf{x}\overset{c}{T}_{00}(t,\mathbf{x)}$ diverges. So, if we are
looking at $\boldsymbol{F}$ giving by Eq.(\ref{pn0}) to have finite energy it
is necessary to impose that
\begin{equation}
\int d^{3}\mathbf{k[}\left\vert u_{+}\left(  \mathbf{k}\right)  \right\vert
^{2}+\left\vert u_{-}\left(  \mathbf{k}\right)  \right\vert ^{2}]<\infty.
\label{pn3}%
\end{equation}
But, what to do with the energy of a monochromatic photon? As already
mentioned in order to give meaning to the energy of a monochromatic photon we
need to go to the formalism of quantum field theory where the classical field
$\boldsymbol{F}(t,\mathbf{x})\mapsto\boldsymbol{\hat{F}}(t,\mathbf{x})$, an
operator valued distribution acting on the appropriated(Fock) Hilbert space
$\mathcal{H}$ for the photon field \ As well know this is done by introducing
creation and destruction operators for the two kind of polarized photons,
i.e., we write \cite{bia}%
\begin{align}
\boldsymbol{\hat{F}}(t,\mathbf{x})  &  =\frac{1}{(2\pi)^{3/2}}\int
d^{3}\mathbf{k}\frac{1}{2\omega\mathbf{_{\mathbf{k}}}}[\mathbf{f}%
_{+}\mathbf{(k)}a\left(  \mathbf{k}\right)  e^{-\mathbf{i}k_{0}%
t+\mathbf{k\bullet x}}+\mathbf{f}_{-}\mathbf{(k)}b^{\dagger}\left(
\mathbf{k}\right)  e^{\mathbf{i}k_{0}t-\mathbf{k\bullet x}}]\nonumber\\
&  +\frac{1}{(2\pi)^{3/2}}\int d^{3}\mathbf{k}\frac{1}{2\omega
\mathbf{_{\mathbf{k}}}}[\mathbf{f}_{-}\mathbf{(k)}b\left(  \mathbf{k}\right)
e^{-\mathbf{i}k_{0}t+\mathbf{k\bullet x}}+\mathbf{f}_{-}\mathbf{(k)}%
a^{\dagger}\left(  \mathbf{k}\right)  e^{\mathbf{i}k_{0}t-\mathbf{k\bullet x}%
}] \label{pn4}%
\end{align}
where $a^{\dagger}\left(  \mathbf{k}\right)  $ and $b^{\dagger}\left(
\mathbf{k}\right)  $ are the creation operators corresponding to left and
right polarizations and $a\left(  \mathbf{k}\right)  $ and $b\left(
\mathbf{k}\right)  $ the respective annihilation operators. We have%
\begin{align}
\lbrack a\left(  \mathbf{k}\right)  ,a^{\dagger}\left(  \mathbf{k}^{\prime
}\right)  ]  &  =\delta(\mathbf{k}-\mathbf{k}^{\prime}),\nonumber\\
\lbrack b\left(  \mathbf{k}\right)  ,b^{\dagger}\left(  \mathbf{k}^{\prime
}\right)  ]  &  =\delta(\mathbf{k}-\mathbf{k}^{\prime}), \label{pn5}%
\end{align}
with all commutators which includes one operator corresponding to left
polarization and other corresponding to right polarization being null.

A one photon state $|\mathrm{ph}\rangle\in\mathcal{H}$ is then written as%
\begin{equation}
|\mathrm{ph}\rangle=\int d^{3}\mathbf{k(}a^{\dagger}(\mathbf{k})f_{+}\left(
\mathbf{k}\right)  +f_{-}\left(  \mathbf{k}\right)  b^{\dagger}(\mathbf{k}%
)|0\rangle\label{pn6}%
\end{equation}
and the the quantum energy density (when Eq.(\ref{pn3}) is normalized to $1$)
of the state $|\mathrm{ph}\rangle$ is given by%
\begin{equation}
\overset{q}{T}_{00}(t,\mathbf{x)=}\langle\mathrm{ph}|\mathbf{:}%
\boldsymbol{\hat{F}}^{\dagger}(t,\mathbf{x})\boldsymbol{\hat{F}}%
(t,\mathbf{x})\mathbf{:}|\mathrm{ph}\rangle\label{pn7}%
\end{equation}
where as usual $\mathbf{:}\boldsymbol{\hat{F}}^{\dagger}(t,\mathbf{x}%
)\boldsymbol{\hat{F}}(t,\mathbf{x})\mathbf{:}$ denotes the normal ordering of
the operators. We have
\begin{equation}
\overset{q}{T}_{00}(t,\mathbf{x)=}\left\vert \boldsymbol{F}_{+}(t,\mathbf{x}%
)\right\vert ^{2}+\left\vert \boldsymbol{F}_{-}(t,\mathbf{x})\right\vert ^{2}
\label{pn8}%
\end{equation}

Since Eq.(\ref{pn8}) shows $\boldsymbol{F}_{+}$ and $\boldsymbol{F}_{-}$
contributes independently in what follows we choose to work with only one
polarization, and we construct following \cite{biab} a localized photon wave
function using the Hertz potential method. describe in previous section. Since
the Hertz potential $\boldsymbol{H}\in\sec%
{\textstyle\bigwedge\nolimits^{2}}
T^{\ast}M$ satisfy the wave equation we write a general free boundary solution
as
\begin{equation}
\boldsymbol{H}(t,\mathbf{x})=\frac{1}{(2\pi)^{4}}\int d^{4}k\delta
(k^{2})\mathbf{h}(k)e^{-\gamma^{5}k\cdot x} \label{pn9}%
\end{equation}
where $\mathbf{h}(k)=\mathbf{h}_{1}+\mathbf{ih}_{2}$ is an arbitrary $2$-form.
Thus, it is
\begin{equation}
\boldsymbol{H}(t,\mathbf{x})=\int d^{3}\mathbf{k}\{\mathbf{h}_{+}%
\mathbf{(k)}e^{-\mathbf{i(}\omega_{\mathbf{k}}t-\mathbf{k\bullet x)}%
}+\mathbf{h}_{\mathbf{-}}^{\ast}\mathbf{(k)}e^{\mathbf{i(}\omega_{\mathbf{k}%
}t-\mathbf{k\bullet x)}}\} \label{pn10}%
\end{equation}
and the positive frequency part of $\boldsymbol{F}_{+}=-d\delta\boldsymbol{H}%
_{+}$ is$\boldsymbol{\ }$%
\begin{equation}
\boldsymbol{F}_{+}(t,\mathbf{x})=\int d^{3}\mathbf{k\{k\times\lbrack
i}\left\vert \mathbf{k}\right\vert \mathbf{\mathbf{h}}_{+}\mathbf{(k)-k\times
\mathbf{h}}_{+}\mathbf{(k)\}}e^{-\mathbf{i}(\omega_{\mathbf{k}}%
t+\mathbf{k\bullet x)}}. \label{pn11}%
\end{equation}
With the choice
\begin{equation}
\mathbf{\mathbf{h}}_{+}\mathbf{(k)=m}\sqrt{l}\left\vert \mathbf{k}\right\vert
^{-\frac{5}{2}}e^{-\frac{\left\vert \mathbf{k}\right\vert l-1}{\left\vert
\mathbf{k}\right\vert l}} \label{pn12}%
\end{equation}
with $\mathbf{m}$ a constant $2$-form it is%
\begin{equation}
\boldsymbol{H}_{+}(t,\mathbf{x})=\mathbf{m}\frac{2\pi^{3/2}}{\mathbf{i}%
r}\left(  e^{-2\sqrt{1+\mathbf{i}\frac{t-r}{l}}}+e^{-2\sqrt{1+\mathbf{i}%
\frac{t+r}{l}}}\right)  . \label{pn13}%
\end{equation}
Thus, $\boldsymbol{F}_{+}$ represents the wave function of a photons having
the form of a spherical shell converging and diverging respectively for
positive and negative values of $t$ and having maximum localizability at
$t=0$. Moreover, $\boldsymbol{H}_{+}$ and also its time derivative at $t=0$
goes as $e^{-\sqrt{2r/l}}$multiplied by some positive power of $r$. This
implies that the photon energy density (Eq.(\ref{pn8})) also decreases
exponentially. We can also show that $\boldsymbol{T}_{0}=T_{0\nu}\gamma^{\nu}$
has an exponential decrease.

\subsection{Can Relativistic Particles be Localized?}

Recall that a single relativistic particle is described by a one particle wave
function obtained from its standard quantum field. As such it satisfy an
hyperbolic equation\footnote{E.g., Klein-Gordon, Dirac and Maxwell equations,
respectively for scalr, spin 1/2 fremions and the photon field.}. As such, a
well result (already quoted \cite{vla}) from the theory of hyperbolic
equations establishes that in solving the Cauchy problem (in a well defined
inertial reference frame) any solution\footnote{And in particular for
solutions that are elements of of a Hilbert space, as required by quantum
theory.} resulting from an initial field configuration having compact support
in space is such that its front and rear will travel with the light speed
$c=1$. For massive particles this immediately leads to a contradiction with
the fundamental assumption of quantum theory that for any time $t$ the wave
function describes the probability amplitude for finding the particle at a
position $\mathbf{x}$ in space\footnote{We could think that for photons such a
problem does not exist, but that is not the case, see below.}.

Indeed, as the particle is supposed to travel at group speed $v_{g}%
=d\omega/d\left\vert \mathbf{k}\right\vert <1$ if its probability amplitude is
described by a wave of compact support in a while the particle and its wave
will uncouples, an absurd.

We can easily prove that localizability for relativistic particles in the
sense of $\delta$-functions is forbidden for this would result in violation of
Einstein causality. \ See e.g., \cite{ticiatti}.

More realistic, as proved in a series of papers by Hegerfeldt (see~\cite{herg}
and references therein) the following fundamental result:

\begin{proposition}
Any solution of a relativistic wave equation in a Hilbert space $\mathcal{H}$
supposed to represent the motion of a particle (and written in such a way that
its Hamiltonian is restricted to be a self-adjoint operator, \emph{ positive}
and \emph{bounded} from below) is such that if a particle is strictly
localized in a bounded region of space $\mathcal{V}$ (as determined in a
inertial reference frame) it happens that for any finite time interval
thereafter the particle localization develops \emph{infinite} tails.
\end{proposition}

This, in particular implies (see, \cite{thaller}) that, e.g., positive
energy-solutions of the Dirac-equation always are noncompact support in space
(have infinity support) which must be the case as we showed above.

Hegerfeldt theorem immediately implies that it is impossible to localize
photons with compact wave functions and the question arises if it is possible
to design localized wave functions for photons which are better localized than
the ones reported, e.g., in \cite{as,nw,sch, jajan}?

Some people think that an yes can be given to that question once one builds
(as it was the case of the above solution) a wave function with an infinity
tail that is exponentially localized at $t=0$. However, for the above solution
and others that already have been reported (see, e.g., \cite{sbz}) the
following criticisms seems to us to invalidate such a claim.

(i) Take into account that the wave function given by Eq.(\ref{pn11}%
)\footnote{And this is the case also for the solution reported in \cite{sbz}%
.}\ exists for all time from $t=-\infty$ up to $t=+\infty$ and thus can only
describe (as it is the case of the PWS that we arrived from the photon concept
following a lightlike worldline in Minkowski spacetime) an \emph{eternally}
propagating photon.

(ii) Note that it seems physically impossible to create a photon described by
a wave function like the one in Eq.(\ref{pn11}) since all photons are produced
in nature in a \emph{finite time, say }$t=T$\ in any emission process and thus
must have a front and a rear.

Hegerfeldt theorem says that such a wave function of compact support (
whatever it may be) immediately develops tails in space, but of course, it is
hard to suppose that it develops tails in \emph{time} for if this was the case
we arrive at the conclusion that the photon wave function start to be create
before it was born!

Anyway, some one may claim that this is only a new non intuitive aspect of
quantum theory.

\subsection{The Quantum Potential}

If we can accept that a photon wave function $\boldsymbol{F}$ ( solution of
the free ME) may be a function that is extended not only in space but also in
time and we moreover think that photons trajectories is a meaningful concept
they may be evaluated as follows:

(i) First from $\boldsymbol{F}$ we evaluate $\mathbf{T}_{0}$ and from our
fundamental assumption $-\partial S=\mathbf{T}_{0}$ we have%
\begin{equation}
S=-\int\mathbf{T}^{0}=\frac{1}{2}\int\boldsymbol{F\gamma}^{0}\boldsymbol{F=}%
\frac{1}{2}\int d\delta\mathbf{\Pi}\boldsymbol{\gamma}^{0}d\delta\mathbf{\Pi}
\label{qp1}%
\end{equation}

(ii) Using Eq.(\ref{qp1}) we can write $\boldsymbol{F}$ as%

\begin{equation}
\boldsymbol{F}=\mathcal{\boldsymbol{F}}e^{\boldsymbol{\gamma}^{5}S}\in%
{\textstyle\bigwedge\nolimits^{2}}
T^{\ast}M\hookrightarrow\mathcal{C\ell(}M,\mathtt{g}) \label{t1}%
\end{equation}
where $\mathcal{\boldsymbol{F}}\in%
{\textstyle\bigwedge\nolimits^{2}}
T^{\ast}M\hookrightarrow\mathcal{C\ell(}M,\mathtt{g})$ is \emph{not} of
course, for a general solution a constant biform. From ME (Eq.(\ref{4n})) we
have\footnote{$\boldsymbol{\partial}\ln\mathcal{F}:=\boldsymbol{\partial
}\mathcal{FF}^{-1}$.}%

\begin{equation}
\boldsymbol{\partial}S\boldsymbol{F}+\boldsymbol{\gamma}^{5}%
\boldsymbol{\partial}\ln\mathcal{\boldsymbol{F}}\boldsymbol{F}=0. \label{t2}%
\end{equation}
Multiplying this equation on the right by $-1/2\boldsymbol{\gamma}%
^{0}\boldsymbol{F}$ and using Eqs.(\ref{0}) and (\ref{2n}) we get%
\begin{equation}
\boldsymbol{\partial}S\boldsymbol{\partial}S+\boldsymbol{\gamma}%
^{5}\boldsymbol{\partial}\ln\mathcal{\boldsymbol{F}}\boldsymbol{\partial}S=0
\label{t3}%
\end{equation}
which means that the \textquotedblleft generalized\textquotedblright\ HJE for
the photon is
\begin{equation}
\boldsymbol{\partial}S\cdot\boldsymbol{\partial}S=-\boldsymbol{\gamma}%
^{5}\boldsymbol{\partial}\ln\mathcal{\boldsymbol{F}}\boldsymbol{\partial
}S=-\langle\boldsymbol{\gamma}^{5}\boldsymbol{\partial}\ln
\mathcal{\boldsymbol{F}}\boldsymbol{\partial}S\rangle_{0} \label{t4}%
\end{equation}
and
\begin{equation}
Q_{\boldsymbol{F}}=\langle\boldsymbol{\gamma}^{5}\boldsymbol{\partial}%
\ln\mathcal{\boldsymbol{F}}\boldsymbol{\partial}S\rangle_{0} \label{14a}%
\end{equation}
and of course it must be the case that the following constraints must be
satisfied%
\begin{equation}
\langle\boldsymbol{\gamma}^{5}\boldsymbol{\partial}\ln\mathcal{\boldsymbol{F}%
}\boldsymbol{\partial}S\rangle_{2}=\langle\boldsymbol{\gamma}^{5}%
\boldsymbol{\partial}\ln\mathcal{\boldsymbol{F}}\boldsymbol{\partial}%
S\rangle_{4}=0. \label{t6}%
\end{equation}

\begin{remark}
Now, for a given solution such that $\boldsymbol{\gamma}^{5}%
\boldsymbol{\partial}\ln\mathcal{\boldsymbol{F}}\boldsymbol{\partial}S\neq0$
in all $M$ we conclude that $\boldsymbol{\partial}S$ is not a lightlike vector
field, so it has an inverse $(\boldsymbol{\partial}S)^{-1}%
=\boldsymbol{\partial}S/\left\vert \boldsymbol{\partial}S\right\vert ^{2}%
$.Thus, multiplying Eq.(\ref{t3}) on the right by $(\boldsymbol{\partial
}S)^{-1}$ we get%
\begin{equation}
\boldsymbol{\partial}S+\boldsymbol{\gamma}^{5}\boldsymbol{\partial}%
\ln\mathcal{\boldsymbol{F}}=0 \label{17T}%
\end{equation}
and we may name
\begin{equation}
Q_{\boldsymbol{F}}=\boldsymbol{\gamma}^{5}\boldsymbol{\partial}\ln
\mathcal{\boldsymbol{F}} \label{18t}%
\end{equation}
the \ \textquotedblleft linearized\textquotedblright\ quantum potential;.
\end{remark}

\begin{remark}
\emph{Eq.(\ref{t4}) }or its linearized version \emph{(Eq.(\ref{17T})) }shows
that the generalized HJE for the photon implies that the \textquotedblleft
would be\textquotedblright\ quantum trajectories are not\ in general lightlike
geodesics of the Minkowski spacetime. At first sight this seems very odd, but
a simple look at a double slit interferometric experiment suggests that indeed
photons are not travelling in lightlike geodesics. Moreover, observe that in
principle $Q_{\boldsymbol{F}}$ can be positive or or null (in some regions,
e.g., besides the double slit screen) implying the existence of timelike and
lightlike world lines. The existence of such non intuitive paths for photons
may eventually lead one to think that eventually such kind of trajectories may
give an answer \emph{(}at least for the case of photons\emph{\footnote{For the
case of electrons the real trajectories must be calculated using the new
generalized Hamilton-Jacobi equation found in \cite{rs1,mrs}.})} for the
pertinent analysis of Chen and Kleinert \emph{\cite{ck}} that Bohmian
trajectories\footnote{According to \cite{bh1993}.} are deficients in view of
the basic quantum mechanics principles.\emph{\footnote{See also the detailed
analysi of Jung \cite{jung} showing the calculated trajectories by
Philippidis, Dewdney and Hiley \cite{pdh} using the nonrelativistic Bohm
theory does not agee with experiment.} }
\end{remark}

\subsection{The Focus Wave Mode Hertz Potential as a Model of a Photon}

Returning to Eq.(\ref{x5}) we can show (see, e.g., \cite{rl1997}) that if we choose%

\begin{equation}
\Xi(\Omega,\beta)=\frac{\pi^{2}}{\mathbf{i}\beta}\exp(-\Omega^{2}z_{0}%
/4\beta),
\end{equation}
we get, assuming $\beta>0$ and $z_{0}>0$, the following Hertz potential
\begin{equation}
\boldsymbol{H}_{fwm}(t,\mathbf{x})=\boldsymbol{\gamma}^{21}\Phi_{fwm}%
(t,\mathbf{x})=\boldsymbol{\gamma}^{21}e^{\mathbf{i}\beta(z+t)}\frac
{\exp\{-\rho^{2}\beta/[z_{0}+\mathbf{i}(z-t)]\}}{4\pi i[z_{0}+\mathbf{i}%
(z-t)]}.
\end{equation}
where the function $\Phi_{fwm}$ is called the \emph{focus wave modus}%
\footnote{It is a special case of Brittingham's \cite{bri} wve focus modus
solutions of the wave equation.}. Function $\Phi_{fwm}$ has very interesting
properties, as discussed in details in$\ $\cite{sbz}. For appropriate choice
of parameters it may describe a wave moving in the positive $z$-direction with
a maximum concentrate in a small region, and that solution is indeed an
improvement over the solution reported above (found in \cite{bia}). Namely,
$\Phi_{fwm}$ is a nondiffracting wave! In this notable paper authors study in
detail the diffraction of $\Phi_{fwm}$ in a modelled double slit experiment.
They show (with appropriate choice of parameters) that after the screen with
the holes the solution of the wave equation generated by the arrival of
$\Phi_{fwm}$ continues to be a function with a well localized maximum
concentrated in a very small region which then hits the detection screen.
Although the classical electromagnetic energy evaluated (trough Eq.(\ref{T1}))
using $\boldsymbol{H}_{fwm}$ diverges, it may be interesting to investigated
how we possibly could renormalize such energy function in order for it to
represent indeed a photon of finite energy. We will return to this issue in
another paper.

\subsection{Some Additional and Pertinent Comments}

(i) Despite the facts presented in the last two sections some authors, e.g.,
Flack and Hiley\ \cite{fh2016} believe that photon trajectories can only be
described by their version of a Bohm like theory for the photon field.
Moreover, they claim that these trajectories can be revealed in weak
measurements of the field momentum. Moreover, they claim that in the double
slit experiment of, e.g., the Toronto experiment \cite{mrf} where their
authors claim to have measured the trajectories of photons \emph{(}although
they based their claim using the nonrelativistic Bohm theory, a nonsequitur in
our opinion) what they indeed measured has been the integral lines of the
Pointing vector evaluated from $\mathbf{T}^{0}$.

Well, we cannot leave out to emphasize that this is what the theory presented
in this paper predicts from the fundamental postulated equation
$\boldsymbol{\partial}S=\frac{1}{2}\boldsymbol{F}\gamma^{0}\boldsymbol{F}$
once $\boldsymbol{F}$ is evaluated as a solution of Eq.(\ref{4n})\emph{ }with
the appropriate boundary and initial conditions, a result distinct from the
Bohmian version of the electromagnetic theory according to \cite{fh2016}.

(ii) Moreover, we must also recall that the formula $\boldsymbol{\partial
}S=\frac{1}{2}\boldsymbol{F}\gamma^{0}\boldsymbol{F}$ fits well the
experimental data on the tunneling of electromagnetic waves in experiments
where the peak of the field configuration seems to go through the barrier with
a \textquotedblleft superluminal speed\textquotedblright\ due to the reshaping
phenomenon that we described above \cite{emig}.

(ii) Finally, we must also mention that the statement in \cite{fh2016} that in
Dirac theory the path of electrons correspond to the integral lines of the
$(0i)$-components of Tetrode energy-mometum tensor is not justified by the
results of \emph{(}see \emph{\cite{rs1,mrs}).} In fact, the study of the
tunneling of electrons through a barrier are well explained (through the
reshaping phenomenon) supposing that electron follows one of the integral
lines of the current $\boldsymbol{J}=e\psi\boldsymbol{\gamma}^{0}\tilde{\psi}$
where $\psi$ is a Dirac-Hestenes spinor field solution of the Dirac-Hestenes
equation for the the conditions of the experiment \cite{dlg}.

\section{Conclusions}

We showed how starting from the photon concept and its relativistic HJE we
immediately get (with a simple hypothesis concerning the form of the photon
canonical momentum) ME satisfied by a null $2$-form field $F$ which is a plane
wave solution (PWS) of ME. Also, it was shown how introducing a potential
$1$-form $\boldsymbol{A}$ such that $\boldsymbol{F}=d\boldsymbol{A}$ we can
see a duality rotation to change in a spatial rotation with the bonus of also
showing how $\mathrm{i}=\sqrt{-1}$ enters Maxwell theory. This permits the
writing of a representative of ME as Schr\"{o}dinger like equation which plays
a key role in answering one of the main questions addressed in this paper,
namely:is there any sense in talking about photon trajectories in de
Broglie-Bohm like theories? To this end we investigate the nature of the
energy-momentum extensor field of the Maxwell field $\mathbf{T}(n)$ in some
special situations which explicitly shows in which sense it seems licit (in
the spirit of de Broglie-Bohm theory) to associate worldlines\ for photons as
being the integral lines of $\mathtt{g}(\mathbf{T}_{0},~)$. We also discuss if
there is any meaning in saying that photons can be say to be localizable due
to the simple fact that there exists free boundary solutions of ME which is
Gaussian concentrated (at least for $t=0$) in the sense of showing an
exponential decay or for the case of nondiffracting solutions like
$\boldsymbol{F}_{fwm}$. Acceptance of such functions as describing photons
implies that we need to accept an odd fact, namely a non localizability in time.

\appendix{}

\section{Maxwell Equations in Vector Formalism}

Let $\{x^{\mu}\}$ be global coordinates for Minkowski spacetime in
Einstein-Lorentz-Poincar\'{e} gauge and $e_{\mu}=\partial/\partial x^{\mu}%
\in\sec TM$,$(\mu,\nu=0,1,2,3)$,.be an orthogonal basis $g(e_{\mu},e_{\nu
})=\eta_{\mu\nu}$ The global field $e_{0}$ determines an \emph{inertial
reference frame}~\cite{rc2016}. Let $\boldsymbol{\gamma}^{\mu}=dx^{\mu}\in\sec%
{\textstyle\bigwedge\nolimits^{1}}
T^{\ast}M\hookrightarrow\sec\mathcal{C}\ell(M,\mathtt{\eta})$ be the dual
basis of $\{e_{\mu}\}$and let $\boldsymbol{\gamma}_{\mu}=\eta_{\mu\nu
}\boldsymbol{\gamma}^{\nu}$ be the reciprocal basis to $\{$
$\boldsymbol{\gamma}^{\mu}\}$, i.e., $\boldsymbol{\gamma}^{\mu}\cdot
\boldsymbol{\gamma}_{\nu}=\delta_{\nu}^{\mu}$. .

The electromagnetic field is represented by a two-form $\boldsymbol{F}\in\sec%
{\textstyle\bigwedge\nolimits^{2}}
T^{\ast}M\hookrightarrow\sec\mathcal{C}\ell(M,\mathtt{\eta})$ We have%

\begin{equation}
\boldsymbol{F}=\frac{1}{2}F^{\mu\nu}\boldsymbol{\gamma}_{\mu}\wedge
\boldsymbol{\gamma}_{\nu}=F^{\mu\nu}\boldsymbol{\gamma}_{\mu}%
\boldsymbol{\gamma},~~~~\ F^{\mu\nu}=\left(
\begin{array}
[c]{cccc}%
0 & -E^{1} & -E^{2} & -E^{3}\\
E^{1} & 0 & -B^{3} & B^{2}\\
E^{2} & B^{3} & 0 & -B^{1}\\
E^{3} & -B^{2} & B^{1} & 0
\end{array}
\right)  , \label{a1}%
\end{equation}
where $(E^{1},E^{2},E^{3})$ and $(B^{1},B^{2},B^{3})$ are respectively the
Cartesian components of the electric and magnetic fields in the reference
frame $e_{0}$.

Let $\boldsymbol{J}\in\sec%
{\textstyle\bigwedge\nolimits^{1}}
T^{\ast}M\hookrightarrow\sec\mathcal{C}\ell(M,\mathtt{\eta})$ be such that%

\begin{equation}
\boldsymbol{J}=J^{\mu}\boldsymbol{\gamma}_{\mu}=\rho\boldsymbol{\gamma}%
_{0}+J^{1}\boldsymbol{\gamma}_{1}+J^{2}\boldsymbol{\gamma}_{2}+J^{3}%
\boldsymbol{\gamma}_{3}, \label{a2}%
\end{equation}
where $\rho$ and $(J^{1},J^{2},J^{3})$ are the Cartesian components of the
charge and (3-dimensional) current densities. Maxwell equation in the language
of differential forms reads%

\begin{equation}
d\boldsymbol{F}=0,\ \delta\boldsymbol{F}=-\boldsymbol{J}, \label{a3}%
\end{equation}
where $d$ is the differential and $\delta$ is the Hodge codifferential operator..

Since $d\boldsymbol{F}$ and $\delta\boldsymbol{F}$ are sections of
$\mathcal{C}\ell(M,\mathtt{\eta})$ we can add the two equations in
Eq.(\ref{a2}) and get%

\[
(d-\delta)\boldsymbol{F}=\boldsymbol{J}.
\]

Now, recalling that $d-\delta=\partial$ is the Dirac operator\footnote{The
operator $\boldsymbol{\partial}$ is not to be confused with the Dirac operator
acting on sections of spin-Clifford bundles. See, \cite{rc2016}.} acting on
sections of $\mathcal{C}\ell(M,\mathtt{\eta})$, and we get%

\begin{equation}
\partial\boldsymbol{F}=\boldsymbol{J} \label{a4}%
\end{equation}
which may now be called \textit{Maxwell equation}, instead of Maxwell equations.

We now write Maxwell equation in $\mathcal{C}\ell^{0}(M,\mathtt{\eta})$, the
even sub-algebra of $\mathcal{C}\ell(M,\mathtt{\eta})$. The typical fiber of
$\mathcal{C}\ell^{0}(M,\mathtt{\eta})$, which is a vector bundle, is
isomorphic to the Pauli algebra (see, e.g., \cite{rc2016}).

We put%

\begin{equation}
\boldsymbol{\sigma}_{i}=\boldsymbol{\gamma}_{i}\boldsymbol{\gamma}%
_{0},\ \mathbf{i}=\boldsymbol{\sigma}_{1}\boldsymbol{\sigma}_{2}%
\boldsymbol{\sigma}_{3}=\boldsymbol{\gamma}_{0}\boldsymbol{\gamma}%
_{1}\boldsymbol{\gamma}_{2}\boldsymbol{\gamma}_{3}=\boldsymbol{\gamma}_{5}.
\label{a5}%
\end{equation}
Recall that $\mathbf{i}$ commutes with bivectors and since $\mathbf{i}^{2}=-1$
it acts like the imaginary unit $\mathrm{i}=\sqrt{-1}$ in $\mathcal{C}\ell
^{0}(M,\mathtt{\eta})$. We now may easily verify that we can write%

\begin{equation}
\boldsymbol{F}=\mathbf{E}+\mathbf{iB} \label{a6}%
\end{equation}
with $\mathbf{E}=E^{i}\boldsymbol{\sigma}_{i}$, $\mathbf{B}=B^{j}%
\boldsymbol{\sigma}_{j}$, $i,j=1,2,3$.

Now, since $\partial=\boldsymbol{\gamma}_{\mu}\partial^{\mu}$ we get
$\partial\boldsymbol{\gamma}_{0}=\partial/\partial x^{0}+\vec{\sigma}%
_{i}\boldsymbol{\gamma}^{i}=\partial/\partial x^{0}-\nabla$. Multiplying
Eq.(\ref{a4}) on the right by $\boldsymbol{\gamma}_{0}$ we have
\begin{equation}
\partial\boldsymbol{\gamma}_{0}\boldsymbol{\gamma}_{0}\boldsymbol{F\gamma}%
_{0}=\mathbf{J}\boldsymbol{\gamma}_{0}, \label{a7}%
\end{equation}
and then%
\begin{equation}
(\partial/\partial x^{0}-\nabla)(-\mathbf{E}+\mathbf{iB})=\rho+\mathbf{J},
\label{a8}%
\end{equation}
where we used $\boldsymbol{\gamma}^{0}\boldsymbol{F\gamma}_{0}=-\mathbf{E}%
+\mathbf{iB}$ and $\mathbf{J:}=J^{i}\boldsymbol{\sigma}_{i}$.

From Eq.(\ref{a8}) we have\footnote{The symbol $\bullet$ denotes the Euclidean
scalar product. Details in~\cite{rc2016} and is not to be confused with the
symbol $\cdot$ that denotes the Minkowskian scalar product.}
\begin{align}
&  -\partial_{0}\mathbf{E}+\mathbf{i}\partial_{0}\mathbf{B}+\nabla
\mathbf{E}-\mathbf{i}\nabla\mathbf{B}=\rho+\mathbf{J}\nonumber\\
&  -\partial_{0}\mathbf{E}+\mathbf{i}\partial_{0}\mathbf{B}+\nabla
\bullet\mathbf{E}+\nabla\wedge\mathbf{E}-\mathbf{i}\nabla\bullet
\mathbf{B}-\mathbf{i}\nabla\wedge\mathbf{B}=\rho+\mathbf{J} \label{a9}%
\end{align}

Now we put (details, e.g., in~\cite{rc2016}) for any $2$-form $\mathbf{A}$ in
$\mathcal{C\ell}(M,\mathtt{g})$ (which can be identified with A euclidean
vector field)%

\begin{equation}
\nabla\times\boldsymbol{A}:=-\mathbf{i}\nabla\wedge\mathbf{A} \label{a10}%
\end{equation}
since the usual vector product between two \textquotedblleft
vectors\textquotedblright\ $\mathbf{A}=A^{i}\boldsymbol{\sigma}_{i}$,
$\mathbf{B}=B^{i}\boldsymbol{\sigma}_{i}$ can be identified with the dual of
the bivector $\mathbf{A}\wedge\mathbf{B}$ through the formula $\mathbf{A}%
\times\mathbf{B}=-\mathbf{i}(\mathbf{A}\wedge\mathbf{B})$. Observe that in
this formalism $\mathbf{A}\times\mathbf{B}$ is a true vector and not the
nonsense pseudo vector of the Gibbs vector calculus. Using Eq.(\ref{a10}) and
equating the terms with the same grade we have%

\begin{equation}%
\begin{array}
[c]{c}%
\nabla\bullet\mathbf{E}=\rho\;;\ \ \ \nabla\times\mathbf{B}-\partial
_{0}\mathbf{E}=\mathbf{J}\;;\\
\\
\nabla\times\mathbf{E}+\partial_{0}\mathbf{B}=0\;;\ \ \ \nabla\bullet
\mathbf{B}=0\;;
\end{array}
\label{a11}%
\end{equation}
which are Maxwell equations in the usual vector notation.

\section{The Symmetrical Energy-Momentum Extensor of the Electromagnetic
Field}

\subsection{$\boldsymbol{\partial}_{n}\cdot\partial\mathbf{T}%
(n)=\boldsymbol{J}\lrcorner\boldsymbol{F}$}

We now introduce the energy momentum extensor of the electromagnetic field and
the energy-momentum 1-forms of stress-energy. Since $\partial\boldsymbol{F}%
=\boldsymbol{J}$ we have $\widetilde{\boldsymbol{\partial F}}=\boldsymbol{J}$.
Multiplying the first equation on the left by $\boldsymbol{F}$ and the second
on the right by $\boldsymbol{F}$ and summing we have:%

\begin{equation}
\frac{1}{2}(\widetilde{\boldsymbol{\partial F}}\boldsymbol{F}%
+\boldsymbol{F\partial F})=\boldsymbol{J}\lrcorner\boldsymbol{F}, \label{a12}%
\end{equation}

Now, let $n$\textbf{ }be a $1$-form field and $\boldsymbol{\gamma}^{\mu}%
\cdot\boldsymbol{\partial}_{n}=\eta^{\mu\nu}\frac{\partial}{\partial n^{\nu}}%
$.Then we can write
\begin{align}
&  \frac{1}{2}(\boldsymbol{F}\widetilde{\boldsymbol{\partial}}\boldsymbol{F}%
+\boldsymbol{F\partial F})\nonumber\\
&  =\frac{1}{2}\left(  \partial_{\mu}\boldsymbol{F\gamma}^{\mu}\boldsymbol{F}%
+\boldsymbol{F\gamma}^{\mu}\partial_{\mu}\boldsymbol{F}\right) \nonumber\\
&  =\frac{1}{2}\boldsymbol{\gamma}^{\mu}\cdot\boldsymbol{\partial}_{n}\left(
\partial_{\mu}\boldsymbol{F}n\boldsymbol{F}+\boldsymbol{F}\partial_{\mu
}n\boldsymbol{F}+\boldsymbol{F}n\partial_{\mu}\boldsymbol{F}\right)
\label{a13}%
\end{align}
where we have used that \ $\boldsymbol{\gamma}^{\mu}\cdot\boldsymbol{\partial
}_{n}\partial_{\mu}n=\eta^{\mu\nu}\frac{\partial}{\partial n^{\nu}}%
\partial_{\mu}n^{\alpha}\boldsymbol{\gamma}^{\alpha}=\eta^{\mu\nu}\delta_{\nu
}^{\alpha}\partial_{\mu}\boldsymbol{\gamma}^{\alpha}=0$. Then we have that%
\begin{equation}
\boldsymbol{\gamma}^{\mu}\cdot\boldsymbol{\partial}_{n}\partial_{\mu}\left(
\frac{1}{2}\boldsymbol{F}n\boldsymbol{F}\right)  =\boldsymbol{J}%
\lrcorner\boldsymbol{F} \label{a14}%
\end{equation}

Eq.(\ref{a14}) means that there exists a differential operator and a $(1,1)$
extensor field\footnote{See \cite{rc2016} for details on the theory of
extensor fields.} defined by

$\boldsymbol{\partial}_{n}\cdot\partial=\boldsymbol{\gamma}^{\mu}%
\cdot\boldsymbol{\partial}_{n}\partial_{\mu}$ and\footnote{If $\boldsymbol{M}$
is a $(1,1)$ extensor field we denote its adjunt by $\boldsymbol{M}^{+}$ and
\ if $n,m$ are $1$-form fields it is: $\boldsymbol{M}(n)\cdot m=n\cdot
\boldsymbol{M}^{+}(m)=\boldsymbol{M}^{+}(m)\cdot n$.}
\begin{gather}
\mathbf{T}^{+}:\sec%
{\textstyle\bigwedge\nolimits^{1}}
T^{\ast}M\hookrightarrow\sec\mathcal{C}\ell(M,\mathtt{\eta})\rightarrow%
{\textstyle\bigwedge\nolimits^{1}}
T^{\ast}M\hookrightarrow\sec\mathcal{C}\ell(M,\mathtt{\eta}),\nonumber\\
\mathbf{T}^{+}(n)=\frac{1}{2}\boldsymbol{F}n\boldsymbol{F} \label{a15}%
\end{gather}

such that
\begin{equation}
\boldsymbol{\partial}_{n}\cdot\partial\mathbf{T}^{+}(n)=\boldsymbol{J}%
\lrcorner\boldsymbol{F}. \label{a16}%
\end{equation}

\subsubsection{$\mathbf{T}(n)=\mathbf{T}^{+}(n)$}

Before going on observe that if $b$ is a $1$-form field it is%
\begin{align}
\mathbf{T}(n)  &  =\boldsymbol{\partial}_{b}n\cdot\mathbf{T}^{+}%
(b)=\boldsymbol{\partial}_{b}n\cdot\frac{1}{2}\boldsymbol{F}b\boldsymbol{F}%
\nonumber\\
&  =\boldsymbol{\partial}_{b}\mathbf{T}^{+}(n)\cdot b=\mathbf{T}^{+}(n),
\label{a17}%
\end{align}
i.e., \textbf{$T$} is symmetric.\medskip

We call the objects \textbf{$T$}$^{\mu}:=$\textbf{$T$}$(\boldsymbol{\gamma
}^{\mu})=T^{\mu\nu}\boldsymbol{\gamma}_{\nu}$ the \emph{energy momentum }%
$1$\emph{-form fields.} It is clear that we have
\begin{equation}
\boldsymbol{\partial\cdot}\mathbf{T}^{\mu}=(\boldsymbol{J}\lrcorner
\boldsymbol{F})\cdot\boldsymbol{\gamma}^{\mu} \label{a18}%
\end{equation}

Of course, if $J=0$ the energy momentum $1$-form fields of the electromagnetic
field is conserved, i.e.,%
\[
\boldsymbol{\partial\cdot}\mathbf{T}^{\mu}=0\Longleftrightarrow\partial_{\mu
}T^{\mu\nu}=0.
\]

We now, define the energy density $u$ and the Pointing vector $\boldsymbol{P}$
by
\begin{align}
\mathbf{T}(\boldsymbol{\gamma}_{0})\boldsymbol{\gamma}_{0}  &  =u+\mathbf{P}%
,\nonumber\\
\boldsymbol{\gamma}_{0}\mathbf{T}(\boldsymbol{\gamma}_{0})  &  =u-\mathbf{P},
\label{a19}%
\end{align}
where
\begin{align}
u  &  =\frac{1}{2}\left(  \mathbf{T}(\boldsymbol{\gamma}_{0}%
)\boldsymbol{\gamma}_{0}+\boldsymbol{\gamma}_{0}\mathbf{T}(\boldsymbol{\gamma
}_{0})\right)  =-\frac{1}{2}\left(  \boldsymbol{F\gamma}_{0}%
\boldsymbol{F\gamma}_{0}+\boldsymbol{\gamma}_{0}\boldsymbol{F\gamma}%
_{0}\boldsymbol{F}\right) \nonumber\\
&  =-\frac{1}{2}\left[  \left(  \mathbf{E}+\mathbf{iB}\right)  \left(
-\mathbf{E}+\mathbf{iB}\right)  +\left(  -\mathbf{E}+\mathbf{iB}\right)
\left(  \mathbf{E}+\mathbf{iB}\right)  \right] \nonumber\\
&  =\frac{1}{2}\left(  \mathbf{E}^{2}+\mathbf{B}^{2}\right)  \label{a20}%
\end{align}
and
\begin{align}
\mathbf{P}  &  =\frac{1}{2}\left(  \mathbf{T}(\boldsymbol{\gamma}%
_{0})\boldsymbol{\gamma}_{0}-\boldsymbol{\gamma}_{0}\mathbf{T}%
(\boldsymbol{\gamma}_{0})\right)  =\frac{1}{2}\left(  \boldsymbol{F\gamma}%
_{0}\boldsymbol{F\gamma}_{0}+\boldsymbol{\gamma}_{0}\boldsymbol{F\gamma}%
_{0}\boldsymbol{F}\right) \nonumber\\
&  =\frac{1}{2}\left[  -\left(  \mathbf{E}+\mathbf{iB}\right)  \left(
-\mathbf{E}+\mathbf{iB}\right)  +\left(  -\mathbf{E}+\mathbf{iB}\right)
\left(  \mathbf{E}+\mathbf{iB}\right)  \right] \nonumber\\
&  =-\mathbf{i}\frac{1}{2}\left(  \mathbf{EB}-\mathbf{BE}\right)
=:\mathbf{E}\times\mathbf{B}. \label{a21}%
\end{align}

From Eq.(\ref{a18}) we can write%
\begin{equation}
\boldsymbol{\partial}\cdot\mathbf{T}_{0}=(J\lrcorner\boldsymbol{F}%
)\cdot\boldsymbol{\gamma}_{0} \label{a22}%
\end{equation}
which we rewrite as%
\begin{align}
\frac{1}{2}(\boldsymbol{\partial}\mathbf{T}_{0}%
+\widetilde{\boldsymbol{\partial}\mathbf{T}_{0}})  &  =\frac{1}{4}\left(
J\boldsymbol{F\gamma}_{0}-\boldsymbol{F}J\boldsymbol{\gamma}_{0}%
+\boldsymbol{\gamma}_{0}J\boldsymbol{F}-\boldsymbol{\gamma}_{0}\boldsymbol{F}%
J\right)  ,\nonumber\\
(\boldsymbol{\partial\gamma}_{0}\boldsymbol{\gamma}_{0}\mathbf{T}%
_{0}+\widetilde{\boldsymbol{\partial\gamma}_{0}\boldsymbol{\gamma}%
_{0}\mathbf{T}_{0}})  &  =\frac{1}{2}[J\boldsymbol{\gamma}_{0}%
\boldsymbol{\gamma}_{0}\boldsymbol{F\gamma}_{0}-\boldsymbol{F}%
J\boldsymbol{\gamma}_{0}+\widetilde{(\boldsymbol{\gamma}_{0}J\boldsymbol{F}%
-\boldsymbol{\gamma}_{0}\boldsymbol{F\gamma}_{0}\boldsymbol{\gamma}_{0}J)}].
\label{a23}%
\end{align}

Now,%
\begin{align}
\boldsymbol{\partial\gamma}_{0}\boldsymbol{\gamma}_{0}\mathbf{T}_{0}  &
=\left(  \partial_{0}-\nabla\right)  (u-\boldsymbol{P})\nonumber\\
&  =\partial_{0}u-\partial_{0}\boldsymbol{P}-\nabla u+\nabla\boldsymbol{P}%
\nonumber\\
&  =\partial_{0}u-\nabla u-\partial_{0}\boldsymbol{P}+\nabla\bullet
\boldsymbol{P}+\mathbf{i}\nabla\times\boldsymbol{P} \label{a24}%
\end{align}
and then%
\begin{equation}
\frac{1}{2}(\boldsymbol{\partial}\mathbf{T}_{0}%
+\widetilde{\boldsymbol{\partial}\mathbf{T}_{0}})=2\left(  \partial
_{0}u+\nabla\bullet\boldsymbol{P}\right)  . \label{a25}%
\end{equation}

Also
\begin{equation}
J\boldsymbol{\gamma}_{0}\boldsymbol{\gamma}_{0}\boldsymbol{F\gamma}%
_{0}-\boldsymbol{F}J\boldsymbol{\gamma}_{0}=-2\left(  \rho\mathbf{E}%
+\mathbf{J}\bullet\mathbf{E}+\mathbf{J}\times\mathbf{B}\right)  \label{a26}%
\end{equation}
and thus
\begin{equation}
\frac{1}{2}[J\boldsymbol{\gamma}_{0}\boldsymbol{\gamma}_{0}\boldsymbol{F\gamma
}_{0}-\boldsymbol{F}J\boldsymbol{\gamma}_{0}+\widetilde{(\boldsymbol{\gamma
}_{0}J\boldsymbol{F}-\boldsymbol{\gamma}_{0}\boldsymbol{F\gamma}%
_{0}\boldsymbol{\gamma}_{0}J)}]=-4\mathbf{J}\bullet\mathbf{E}. \label{a27}%
\end{equation}
Using this results we can rewrite Eq.(\ref{a22}) as%
\begin{equation}
\frac{\partial u}{\partial T}=-(\nabla\bullet\boldsymbol{P}+\mathbf{J}%
\bullet\mathbf{E}), \label{a28}%
\end{equation}
a result which is known as Poynting theorem.

\subsection{Explicit Form of the Components $T^{\mu\nu}$}

Since $\boldsymbol{F\gamma}^{\mu}\boldsymbol{F}\in\sec%
{\textstyle\bigwedge\nolimits^{1}}
T^{\ast}M\hookrightarrow\sec\mathcal{C}\ell(M,\mathtt{\eta})$\ we can write%

\begin{equation}
T^{\mu\nu}=-\frac{1}{2}(\boldsymbol{F\gamma}^{\mu}\boldsymbol{F}%
)\cdot\boldsymbol{\gamma}^{\nu}=-\frac{1}{2}\langle\boldsymbol{F\gamma}^{\mu
}\boldsymbol{F\gamma}^{\nu}\rangle_{0} \label{a29}%
\end{equation}

Since $\boldsymbol{\gamma}^{\mu}\lrcorner\boldsymbol{F}={\frac{1}{2}%
}(\boldsymbol{\gamma}^{\mu}\boldsymbol{F}-\boldsymbol{F\gamma}^{\mu
})=-\boldsymbol{F}\llcorner\boldsymbol{\gamma}^{\mu}$, we have
\begin{align}
\frac{1}{2}\langle\boldsymbol{F\gamma}^{\mu}\boldsymbol{F\gamma}^{\nu}%
\rangle_{0}  &  =-\langle(\boldsymbol{\gamma}^{\mu}\lrcorner\boldsymbol{F}%
)\boldsymbol{F\gamma}^{\nu}\rangle_{0}+\frac{1}{2}\langle\boldsymbol{\gamma
}^{\mu}\boldsymbol{F}^{2}\boldsymbol{\gamma}^{\nu}\rangle_{0}\nonumber\\
&  =+(\boldsymbol{\gamma}^{\mu}\lrcorner\boldsymbol{F})\cdot
(\boldsymbol{\gamma}^{\nu}\lrcorner\boldsymbol{F})+\frac{1}{2}(\boldsymbol{F}%
\lrcorner\boldsymbol{F})\boldsymbol{\gamma}^{\mu}\cdot\boldsymbol{\gamma}%
^{\nu}\nonumber\\
&  =\frac{1}{4}\boldsymbol{F}^{\mu\iota}\boldsymbol{F}_{\cdot l}^{\nu\cdot
}-\frac{1}{2}(\boldsymbol{F}\cdot\boldsymbol{F})\boldsymbol{\gamma}^{\mu}%
\cdot\boldsymbol{\gamma}^{\nu}\nonumber\\
&  =\frac{1}{4}\boldsymbol{F}^{\mu\iota}\boldsymbol{F}_{\cdot l}^{\nu\cdot
}-\frac{1}{8}\boldsymbol{F}_{\alpha\beta}\boldsymbol{F}^{\alpha\beta}\eta
^{\mu\nu}. \label{a30}%
\end{align}

Thus,%
\begin{equation}
T^{\mu\nu}=\frac{1}{4}\left(  -\boldsymbol{F}^{\mu\iota}\boldsymbol{F}%
_{\cdot\iota}^{\nu\cdot}+\frac{1}{4}\boldsymbol{F}_{\alpha\beta}%
\boldsymbol{F}^{\alpha\beta}\eta^{\mu\nu}\right)  \label{a31}%
\end{equation}

\subsection{Angular Momentum Extensor}

We now define the density of \textit{angular momentum extensor}. Take a point
event $\mathfrak{o\in}M$ such that its coordinates are $(0,0,0,0)$ and define
1-form $\boldsymbol{x}=x^{\mu}\boldsymbol{\gamma}_{\mu}=x_{\mu}%
\boldsymbol{\gamma}^{\mu}$. The \textit{angular momentum extensor is the
mapping}%
\begin{align}
\boldsymbol{M}^{\dagger}  &  :\sec%
{\textstyle\bigwedge\nolimits^{1}}
T^{\ast}M\hookrightarrow\sec\mathcal{C}\ell(M,\mathtt{\eta})\rightarrow%
{\textstyle\bigwedge\nolimits^{2}}
T^{\ast}M\hookrightarrow\sec\mathcal{C}\ell(M,\mathtt{\eta}),\nonumber\\
\boldsymbol{M}^{\boldsymbol{\dagger}}(n)  &  =\mathbf{T}(n)\wedge
\boldsymbol{x}=\frac{1}{2}(\mathbf{T}(n)\boldsymbol{x}-\boldsymbol{x}%
\mathbf{T}(n)) \label{a32}%
\end{align}
In particular we define the angular momentum $2$-form fields by%

\begin{equation}
\boldsymbol{M}_{\mu}^{\boldsymbol{\dagger}}=\boldsymbol{M}%
^{\boldsymbol{\dagger}}(\boldsymbol{\gamma}_{\mu})=\mathbf{T}_{\mu}%
\wedge\boldsymbol{x}=\frac{1}{2}(x_{\alpha}T_{\mu\nu}-x_{\nu}T_{\alpha\mu
})\boldsymbol{\gamma}^{\nu}\wedge\boldsymbol{\gamma}^{\alpha}. \label{a33}%
\end{equation}
We immediately get that%
\begin{equation}
\boldsymbol{\partial}_{n}\cdot\boldsymbol{\partial M}^{\boldsymbol{\dagger}%
}(n)=-\boldsymbol{x}\wedge(J\lrcorner\boldsymbol{F}) \label{A34}%
\end{equation}
and in particular it is%

\begin{equation}
\partial^{\mu}\boldsymbol{M}_{\mu}^{\boldsymbol{\dagger}}=-\frac{1}%
{2}(x^{\alpha}J_{\mu}\boldsymbol{F}^{\mu\nu}-x^{\nu}J_{\mu}\boldsymbol{F}%
^{\mu\alpha})\boldsymbol{\gamma}_{\alpha}\wedge\boldsymbol{\gamma}_{\nu}.
\label{a35}%
\end{equation}

\begin{remark}
It is important to emphasize here that without the Lagrangian formalism it is
not possible to identify the spin extensor of the electromagnetic field
\cite{rc2016}.
\end{remark}

\subsection{Poincar\'{e} Invariants}

The Poincar\'{e} \textit{invariants} of the electromagnetic field
$\boldsymbol{F}$ are $\boldsymbol{F}\lrcorner\boldsymbol{F}$ and
$\boldsymbol{F}\wedge\boldsymbol{F}$ and
\[
\boldsymbol{F}^{2}=\boldsymbol{F}\lrcorner\boldsymbol{F}+\boldsymbol{F}%
\wedge\boldsymbol{F};
\]%
\begin{equation}
\boldsymbol{F}\lrcorner\boldsymbol{F}=-\frac{1}{2}\boldsymbol{F}^{\mu\nu
}\boldsymbol{F}_{\mu\nu};\qquad\boldsymbol{F}\wedge\boldsymbol{F}=\frac{1}%
{4}\boldsymbol{F}_{\mu\nu}\boldsymbol{F}_{\alpha\beta}\varepsilon^{\mu
\nu\alpha\beta}\boldsymbol{\gamma}^{0}\boldsymbol{\gamma}^{1}%
\boldsymbol{\gamma}^{2}\boldsymbol{\gamma}^{3}=\frac{1}{4}\boldsymbol{\gamma
}^{5}\varepsilon^{\mu\nu\alpha\beta}\boldsymbol{F}_{\mu\nu}\boldsymbol{F}%
_{\alpha\beta}. \label{a36}%
\end{equation}

Writing as before $\boldsymbol{F}=\mathbf{E}+\mathbf{iB}$ we have%

\begin{equation}
\boldsymbol{F}^{2}=(\mathbf{E}^{2}-\mathbf{B}^{2})-2\mathbf{iE}\times
\mathbf{B}=\boldsymbol{F}\lrcorner\boldsymbol{F}+\boldsymbol{F}\wedge
\boldsymbol{F}. \label{a37}%
\end{equation}

\subsection{The Canonical Energy-Momentum Extensor of the Electromagnetic
Field}

The action for the free Maxwell field $\boldsymbol{F}=d\boldsymbol{A}$,
$\boldsymbol{A}\in\sec%
{\textstyle\bigwedge\nolimits^{1}}
T^{\ast}M\hookrightarrow\sec\mathcal{C}\ell(M,\mathtt{\eta})$ is%
\begin{align}
S  &  =\int d^{4}x\mathfrak{L}(\boldsymbol{A},\boldsymbol{\partial}%
\wedge\boldsymbol{A}),\nonumber\\
\mathfrak{L}(\boldsymbol{A},\boldsymbol{\partial}\wedge\boldsymbol{A})  &
=-\frac{1}{2}(\boldsymbol{\partial}\wedge\boldsymbol{A})\cdot
(\boldsymbol{\partial}\wedge\boldsymbol{A}) \label{a38}%
\end{align}
and the canonical energy-momentum tensor is \cite{rc2016}%
\begin{equation}
\mathbf{T}_{c}^{\dagger}(n)=\mathbf{T}^{\dagger}(n)-(n\lrcorner\boldsymbol{F}%
)\cdot\boldsymbol{\partial A}. \label{a39}%
\end{equation}

The canonical formalism naturally gives a conserved total angular momentum
$(1,2)$-extensor field for any field theory. For our case, we have that
\begin{gather}
\boldsymbol{J}^{\dagger}:\sec%
{\textstyle\bigwedge\nolimits^{1}}
T^{\ast}M\hookrightarrow\sec\mathcal{C}\ell(M,\mathtt{\eta})\rightarrow%
{\textstyle\bigwedge\nolimits^{2}}
T^{\ast}M\hookrightarrow\sec\mathcal{C}\ell(M,\mathtt{\eta}),\nonumber\\
\boldsymbol{J}^{\dagger}(n)=\mathbf{T}_{c}^{\dagger}(n)\wedge\boldsymbol{x+S}%
^{\dagger}(n), \label{a40}%
\end{gather}
where \footnote{With $\times$ in Eq.(\ref{A41}) denoting the commutator
product in the Clifford bundle, i.e., for $K,L\in\sec\mathcal{C\ell
}(M,\mathtt{\eta})$ it is $K\times L=1/2(KL-LK)$.}
\begin{align}
\boldsymbol{S}^{\dagger}(n)  &  =\langle\boldsymbol{A}\times\partial
_{\boldsymbol{\partial\wedge A}}\mathfrak{L}(\boldsymbol{A}%
,\boldsymbol{\partial}\wedge\boldsymbol{A})n\rangle_{2}\nonumber\\
&  =(n\lrcorner\boldsymbol{F})\wedge\boldsymbol{A}=n\lrcorner(\boldsymbol{A}%
\wedge\boldsymbol{F})-(n\cdot\boldsymbol{A})\boldsymbol{F} \label{A41}%
\end{align}
is the spin extensor of the electromagnetic field which is a gauge dependent
quantity.\cite{devries,gr2012}.

\end{document}